\documentclass[12pt]{article}
\usepackage{amsmath}
\usepackage{graphicx,psfrag,epsf}
\usepackage{enumerate}
\usepackage{url}
\usepackage{multirow}
\usepackage{amsfonts}
\usepackage{amssymb}
\usepackage[page]{appendix}
\usepackage{algorithm}
\usepackage{algpseudocode}

\newcommand{\blind}{0}
\newtheorem{theorem}{Theorem}

\newtheorem{proposition}[theorem]{Proposition}

\newenvironment{proof}[1][Proof]{\textbf{#1.} }{\ \rule{0.5em}{0.5em}}

\begin{document}

\def\spacingset#1{\renewcommand{\baselinestretch}%
{#1}\small\normalsize} \spacingset{1}


\if0\blind
{ \title{\bf Understanding fluctuations through Multivariate Circulant Singular Spectrum Analysis}
  \author{Juan B\'{o}galo \thanks{
    The authors acknowledge financial help provided by the Spanish Ministry of Science and Innovation, 
    contract grants \textit{MINECO/FEDER PID2019-107161GB-C32 and PID2019-108079GB-C22}}\hspace{.2cm}\\
\small Universidad Aut\'{o}noma de Madrid\\
\small SPAIN \\
\and
Pilar Poncela \\
\small Universidad Aut\'{o}noma de Madrid\\
\small SPAIN 
\and 
Eva Senra \\
\small Universidad de Alcal\'{a}\\
\small SPAIN}
  \maketitle
} \fi

\if1\blind
{
  \bigskip
  \bigskip
  \bigskip
  \begin{center}
    {\LARGE\bf Understanding fluctuations through Multivariate Circulant Singular Spectrum Analysis}
  \end{center}
  \medskip
} \fi

\begin{abstract}
We introduce Multivariate Circulant Singular Spectrum Analysis (M-CiSSA) to provide a comprehensive framework to analyze fluctuations, extracting the underlying components of a set of time series, disentangling their sources of variation and assessing their relative phase or cyclical position at each frequency. Our novel method is non-parametric and can be applied to series out of phase, highly nonlinear and modulated both in frequency and amplitude. We prove a uniqueness theorem that in the case of common information and without the need of fitting a factor model, allows us to identify common sources of variation. This technique can be quite useful in several fields such as climatology, biometrics, engineering or economics among others.
We show the performance of M-CiSSA through a synthetic example of latent signals modulated both in amplitude and frequency and through the real data analysis of energy prices to understand the main drivers and co-movements of primary energy commodity prices at various frequencies that are key to assess energy policy at different time horizons. 
\end{abstract}

\noindent%
\vfill

\newpage
\spacingset{1.45} 
\section{Introduction}
\label{sec:intro}

Signal extraction is an old problem across various disciplines, but typical decomposition procedures differ depending on the field. 
For instance, engineers work with amplitude and frequency modulated (AM-FM) signals.
They usually deal with high frequencies and use different methods to extract possibly highly non-linear fluctuations, such as those based on the Hilbert Transform (Huang et al., 1998; Gianfelici et al., 2007; and Biagetti et al., 2015).
However, these methods were not designed to work with low frequencies.
On the other hand, economists need to extract long and medium-term information, like the trend and cycle, free of seasonality, for policy analysis. 
In general, they use parametric models either based on ARIMA formulations (see, e.g. Hillmer and Tiao, 1982) or in the state space framework (see e.g. Harvey, 1989) that may not work well with short or highly nonlinear time series. 

Singular Spectrum Analysis (SSA) is another alternative for signal extraction based on subspace algorithms (see, for instance, the surveys by Ghil et al., 2002, or Golyandina and Zhigljavsky, 2013). After choosing a window length $L$, SSA  builds a related trajectory matrix by putting together lagged pieces of the original time series and performs its Singular Value Decomposition. Different alternatives of SSA, such as Basic and Toeplitz SSA, need to identify the frequencies associated with the estimated components after they have been extracted. 
Within this framework Circulant Singular Spectrum Analysis (CiSSA), is a nonparametric procedure developed in Bógalo et al. (2021) for univariate time series, that allows for an automated matching between the extracted signals and their frequency of oscillation.

The multivariate extension of Singular Spectrum Analysis, M-SSA, (Broomhead and King, 1986b) appears simultaneously with univariate SSA (Broomhead and King, 1986a; Fraedrich, 1986). 
M-SSA, as is the case with its univariate counterpart, also needs to identify the frequencies of the reconstructed components and, as Plaut and Vautard (1994) observe, is able to extract patterns both along time and across time series. While originally only applied to climatology, since the work of Ghil et al. (2002) who compared different versions, the application of this technique has been extended to a wide range of disciplines: biometrics (see, e.g., Lee et al., 2014), seismic activity (see, e.g., Cheng et al., 2019), geolocalization (see, e.g., Gruszczynska et al., 2017), business cycles and economics (see, e.g., Carvalho (de) and Rua, 2017; Hassani et al., 2013; Silva et al., 2018) and medical diagnostic (see, e.g., Jain et al., 2020).

Multivariate Circulant Singular Spectrum Analysis (M-CiSSA) is a novel unified framework for multivariate signal extraction. 
It builds a new trajectory matrix, different to previous M-SSA, and a related block circulant matrix of second order moments that allows us to compute the cross-spectral density matrices at different frequencies in a straightforward way. 
It is based on the properties of the eigen-structure of a block circulant matrix related to the second order moments of the vector of time series and their lags.
Also, further diagonalization of these matrices enables us to obtain the eigenvectors and eigenvalues that provide the principal components and their contribution to the total variability within a specific frequency. 
Both aspects, block frequency identification and decomposition within a block, are new contributions in this setup. As a consequence, we can prove a uniqueness theorem  that states the reconstruction of the univariate components by the sum of the multivariate subcomponents per frequency. 

M-CiSSA offers several advantages. Firstly, eigenvalues and eigenvectors of each of these blocks resulting from the block diagonalization contain all the variability of the corresponding frequencies. This enables us to understand co-movements at different frequencies and even the cyclical position among variables. Secondly, the uniqueness theorem is new to SSA and is very useful to further understand the formation of the individual cycles in terms of the multivariate common drivers. Thirdly, the solution proposed can be applied to any range of frequencies and to very different types of time series.

All in all, our approach provides a comprehensive framework for jointly analyzing  fluctuations, solving, in a unified way, the problem of extracting the underlying components of a set of time series, disentangling their sources of variation and assessing their relative cyclical position at each frequency. 
In the particular case of common information and without the need for fitting a factor model, M-CiSSA also identifies common sources of variation and assigns them to a particular frequency. It can also serve as a tool for denoising the latent signals.

We illustrate the performance of this technique through two examples. First, we use a synthetic example of two series, each of them generated as the sum of various latent signals including amplitude and modulated ones. We show that M-CiSSA is able to clearly separate the underlying signals. We complete our illustrations with the analysis of a real data set of Primary Commodity Energy Prices in order to characterize the main latent signals (trends, cycles) driving their fluctuations. We found co-movements in the long and medium term for all series except US natural gas, and identified the additional decoupling of coals and Japanese natural gas only at the medium-term cyclical frequency. Finally, we have discovered which prices can be held or sustained by forces outside the common evolution of the markets.

The remainder of the paper is structured as follows. Section 2 presents the proposed  methodology. Section 3 shows the two data applications.  Finally, we draw our conclusions in section 4.

\section{M-CiSSA}
In this section we will present our new proposal that generalizes the univariate CiSSA (Bógalo et al. 2021). The theory behind CiSSA starts, as any other SSA procedure, with the transformation of the original series into a related trajectory matrix, its decomposition into elementary units, and further reconstruction to regain the original time series dimensions.
The novelty is that the decomposition is based on a circulant second order matrix that guarantees an explicit expression for the eigenvalues and eigenvectors that relates them to frequencies.
This perfect match between frequencies and the eigen-structure of the matrix allows us for an automated identification of the signals in the time series.
The multivariate M-CiSSA follows a similar approach that substitutes the circulant matrix by a block circulant matrix that allows us to block diagonalize and match each block with a frequency. 
While each block contains all the information related to the multivariate frequency, further diagonalization within the block will identify the main drivers that explain the fluctuations.

 For a better understanding of the new proposal, we will first review the univariate algorithm. Afterwards, we will introduce the steps of the multivariate version highlighting the novelties and generalizations that were needed compared to the univariate case. Finally, we will present the auxiliary results that are the pillars of our new proposal. 


\subsection{Univariate CiSSA}

In what follows, let $\left\{ x_{t}\right\}_{t=1} ^{T}$ be a realization of a stochastic process   $\left\{ x_{t}\right\}$  $t \in \cal T$ and  $\mathbf{x}=(x_{1},...,x_{T})^{\prime}$, where the prime denotes transpose and $L$ a positive integer, called the
window length, such that $1<L<T/2$.\footnote{See that we use the same notation for the stochastic process and for the observed time series. When needed, it will be explicitly clarified in the main text which one we are referring to.} 
The algorithm works in four steps\footnote{For additional details, see Bógalo et al. (2021).}: 

\bigskip
\textbf{1st step: Embedding.}
Select the window length $L$ and convert the univariate time series into a matrix by defining a trajectory matrix $\mathbf{X}$, $N=T-L+1$, as follows
\begin{equation}
\mathbf{X=}\left( \mathbf{x}_{1}|...|\mathbf{x}_{N}\right) =\left( 
\begin{array}{ccccc}
x_{1} & x_{2} & x_{3} & ... & x_{N} \\ 
x_{2} & x_{3} & x_{4} & ... & x_{N+1} \\ 
\vdots & \vdots & \vdots & \vdots & \vdots \\ 
x_{L} & x_{L+1} & x_{L+2} & ... & x_{T}%
\end{array}%
\right) \label{trajectohe ry}
\end{equation}%
where $\mathbf{x}_{j}=(x_{j},...,x_{j+L-1})^{\prime }$ indicates the $L \times 1$ vector
with origin at time $j$. 

\bigskip
\textbf{2nd step: Decomposition}
Compute the second moments  
\begin{equation}
\widehat{\gamma }_{m}=\frac{1}{T-m}\sum_{t=1}^{T-m}x_{t}x_{t+m}\: \label{gamma}
\end{equation}
and define  
\begin{equation}
\widehat{c}_{m}=\frac{L-m}{L}\widehat{\gamma }_{m}+\frac{m}{L}\widehat{\gamma }_{L-m},\qquad m=0,1,...,L-1\:. \label{c_m}
\end{equation}
that are the elements of the circulant matrix 
\begin{equation}
\mathbf{S}_{C} =\left( 
\begin{array}{ccccc}
\widehat{c}_{0} & \widehat{c}_{1} & \widehat{c}_{2} & ... & \widehat{c}_{L-1} \\ 
\widehat{c}_{L-1} & \widehat{c}_{0} & \widehat{c}_{1} & ... & \widehat{c}_{L-2} \\ 
\vdots & \vdots & \vdots & \vdots & \vdots \\ 
\widehat{c}_{1} & \widehat{c}_{2} & \widehat{c}_{3} & ... & \widehat{c}_{0}%
\end{array}%
\right) \label{trajectohe ry}
\end{equation}%
and are also needed to compute its eigenvalues\begin{equation*}
\widehat{\lambda}_{L,k}=\sum_{m=0}^{L-1}\widehat{c}_{m}\exp \left( i2\pi m\frac{k-1}{L}\right)
\end{equation*}

The associated eigenvectors are given by   $k=1,...,L$ 
\begin{equation}
\mathbf{u}_{k}=L^{-1/2}(u_{k,1,}...,u_{k,L})^{\prime } \label{u_k}
\end{equation}
where $u_{k,j}=\exp \left( -i2\pi (j-1)\frac{k-1}{L}\right)$.

Form the elementary matrices of rank 1, $\mathbf{X}_{k}=\mathbf{u}_{k}\mathbf{u}_{k}^*\mathbf{X}$, where the superindex $*$ denotes the conjugate transpose.

\bigskip
\textbf{3rd step: Grouping.}
Based on the following relationship between the eigenvalues and the spectral density function of the data $f$ (see Lancaster, 1969) 
\begin{equation}
\lambda _{L,k}=f\left( \frac{k-1}{L}\right), \label{auto}
\end{equation}%
associate the $k$-th eigenvalue and corresponding eigenvector to the frequency $w_k=\frac{k-1}{L},k=1,...,L.$
Group the elementary matrices $\mathbf{X}_{k}$ into $G$ disjoint groups. For each group $j=1,...,G$, we can define the matrix $\mathbf{X}_{I_{j}}=\sum_{k\in I_j} \mathbf{X}%
_k$ summing all the elementary matrices within this group.

\bigskip
\textbf{4th step: Reconstruction.}
Convert each matrix from step 3 into a time series of the original dimension, denoted as $\widetilde{\mathbf{x}}^{(j)}=(%
\widetilde{x}_{1}^{(j)},...,\widetilde{x}_{T}^{(j)})^{\prime }$ by diagonal
averaging. 
Denoting by $\widetilde{x}_{r,s}$ the elements of the matrix $\mathbf{X}_{I_{j}}$, the reconstruction is done by averaging the elements of this matrix over its antidiagonals,
\begin{equation*}
\widetilde{x}_{t}^{(j)}=\text{H}\left(\mathbf{X}_{I_j}\right) = \left\{ 
\begin{array}{l}
\frac{1}{t}\sum_{i=1}^{t}\widetilde{x}_{i,t-i+1},\qquad 1\leq t<L \\ 
\frac{1}{L}\sum_{i=1}^{L}\widetilde{x}_{i,t-i+1},\qquad L\leq t\leq N \\ 
\frac{1}{T-t+1}\sum_{i=L-N+1}^{T-N+1}\widetilde{x}_{i,t-i+1},\qquad N<t\leq T%
\end{array}%
\right .
\end{equation*}
and constitutes the extracted signals associated to a particular frequency or range of frequencies.

Note that with CiSSA the analyst just needs to choose the desired frequencies beforehand to obtain the related signals.
This is different from previous attempts to automate SSA where the components are first estimated and in a second step associated to a frequency, see Ghil and Mo (1991), Vautard et al. (1992), Alexandrov and Golyandina (2005), Alexandrov (2009), Alonso and Salgado (2008), Bilancia and Campobasso (2010), Arteche and García-Enríquez (2017), Carvalho (de) and Rúa (2017) and recently, Golyandina and Zhornikova (2023), among others.

\subsection{Multivariate CiSSA}
The goal we pursue with M-CiSSA is more ambitious as it aims to understand the formation of fluctuations by frequency, uncovering their commonalities and specificities, and also the relative cyclical position of the signals.
The M-CiSSA solution is based on the block diagonalizaton of a block circulant matrix related to the second moments of the data. 
Because of the properties of block circulant matrices, there is an exact match that associates a block with a specific known frequency.  
Further diagonalization within each block explains the cross-section variability by frequency and, therefore, allows us to  understand how series are related at each frequency. 
A new insight of the multivariate problem is the so-called multichannel analysis or the knowledge of how each series contributes to the formation of the signal. 
We will extract the unobserved signals from several channels or time series in contrast to the univariate analysis where the trend and different cycles were only fed by one series or channel. When the different time series are measures of a particular variable in different points of a grid, this gives a space-time decomposition. The eigenvalue problem to be solved is similar to that of the univariate case but replacing each data point by a vector of $M$ time series. This rises another issue and is how each channel or univariate time series contributes to the formation of the different unobserved signals. In this sense, each element of the eigenvectors in univariate analysis is now replaced by a segment or subvector of size $M$. Putting together the elements coming from each time series, we can find the contribution of each one of them to the formation of the unobserved signals. We will make explicit in our proposal how to identify the "pieces" of eigenvectors corresponding to each channel or time series. 

So, in what follows, we will first introduce the new trajectory matrix; then, we will present the four steps of the M-CiSSA procedure and, finally, we will justify the new algorithm. 



We start by defining the new trajectory matrix. Let $ \mathbf{x}_t=\left(x_t^{\left(1\right)},\cdots,x_t^{\left(M\right)}\right)'$, be an $M$-dimensional stationary and, for the sake of simplicity, zero mean stochastic process $t \in \cal T$ with autocovariance matrices $ \mathbf{\Gamma}_k= E[\mathbf{x}_{t+k}\mathbf{x}_t^{\prime}], k=0,\pm1,\cdots,\pm\left(L-1\right)$. Consider now its realization of length $T$.
Given a window length $L<T/2$ we construct a trajectory matrix, of dimensions $LM\times N$ with $ N=T-L+1$ as,

\begin{equation}
\mathbf{X}=\left(\begin{matrix}\mathbf{x}_1&\mathbf{x}_2&\cdots&\mathbf{x}_N\\\mathbf{x}_2&\mathbf{x}_3&\cdots&\mathbf{x}_{N+1}\\\vdots&\vdots&\vdots&\vdots\\\mathbf{x}_L&\mathbf{x}_{L+1}&\cdots&\mathbf{x}_T\\\end{matrix}\right)=
\left(\begin{matrix}x_1^{\left(1\right)}&x_2^{\left(1\right)}&\cdots&x_N^{\left(1\right)}\\
\vdots&\vdots&\vdots&\vdots\\
x_1^{\left(M\right)}&x_2^{\left(M\right)}&\cdots&x_N^{\left(M\right)}\\
\vdots&\vdots&\vdots&\vdots\\x_L^{\left(1\right)}&x_{L+1}^{\left(1\right)}&\cdots&x_T^{\left(1\right)}\\\vdots&\vdots&\vdots&\vdots\\x_L^{\left(M\right)}&x_{L+1}^{\left(M\right)}&\cdots&x_T^{\left(M\right)}.\\\end{matrix}\right)
\label{eq_trajectory}
\end{equation}

Notice that this construction of the trajectory matrix differs from that of other SSA multivariate algorithms\footnote{The trajectory matrix in other SSA multivariate algorithms stacks the univariate trajectory matrices instead of considering each value of the matrix as a vector of time series.}. 
The consideration of this alternative trajectory matrix is a relevant issue, since its matrix of second order moments is block Toeplitz and will lay the foundation for the construction of the related block circulant matrix.

\subsection{The four steps in M-CiSSA}
 The four steps of the new M-CiSSA algorithm present several novelties. First, on the embedding step we will use the new trajectory matrix that we have previously defined. This will allow us to define the blocks of the elementary matrices associated with each time series in the decomposition step. The diagonalization will be done in two steps: block diagonalization, first, followed by a further diagonalization within each block. Putting together the pieces of eigenvectors corresponding to the same time series is easy in our case as the element of the subvector corresponding to series $i$ occupies always the same position $i$ within the subvector and allows us to easily identify the contribution of each channel or time series to the formation of the unobserved components.  As with univariate CiSSA, the grouping will be automated and we extend the results of the univariate case, where we match frequencies with eigenstructure, to the multivariate setup. 
 As an auxiliary result, we need to show that the new proposed circulant matrices are asymptotically equivalent to the traditional variance-covariance matrices. The reconstruction step is done as usual in SSA algorithms. In what follows we are going to, first, present the 4 steps of the algorithm and, afterwards, to show the results that allow us to introduce the aforementioned novelties 
 
The four steps of the M-CiSSA algorithm are:

\textbf{1st step: Embedding.}

Form the big trajectory matrix $\mathbf{X}$ as in (\ref{eq_trajectory}).
 
\bigskip
\textbf{2nd step: Double decomposition.}

Compute the sample second moment matrices $\hat{\mathbf{\Gamma}}_k=\frac{1}{T-k}\sum_{t=1}^{T-k}\mathbf{x}_{t+k}\mathbf{x}_t^\prime$, $k=1,\cdots,L-1$ and from them, the linear combinations
\begin{equation}
{\hat{\mathbf{\Omega}}}_k=\frac{k}{L}\hat{\mathbf{\Gamma}}_{L-k}+\frac{L-k}{L}\hat{\mathbf{\Gamma}}_{-k}, \; k=0,\cdots,L-1.
\label{omega_tilde}
\end{equation}
Build the block circulant matrix $\mathbf{S}_\mathbf{C}$ given by
\begin{equation*}
\mathbf{S}_\mathbf{C}=\left(\begin{matrix}{\hat{\mathbf{\Omega}}}_0&{\hat{\mathbf{\Omega}}}_1&\cdots&{\hat{\mathbf{\Omega}}}_{L-1}\\{\hat{\mathbf{\Omega}}}_{L-1}&{\hat{\mathbf{\Omega}}}_0&\ddots&\vdots\\\vdots&\ddots&\ddots&{\hat{\mathbf{\Omega}}}_1\\{\hat{\mathbf{\Omega}}}_1&\cdots&{\hat{\mathbf{\Omega}}}_{L-1}&{\hat{\mathbf{\Omega}}}_0\\\end{matrix}\right)
\label{S_C}
\end{equation*}

Diagonalize the $LM \times LM$ matrix $\mathbf{S}_\mathbf{C}$ in two steps.
\begin{enumerate}
\item Block diagonalization by frequency.

The block spectral decomposition of $\mathbf{S}_\mathbf{C}$ is given by ${{\mathbf{S}}_{C}}=\left( {{\mathbf{U}}_{L}}\otimes {{\mathbf{I}}_{M}} \right)\mathbf{\hat{F}}{{\left( {{\mathbf{U}}_{L}}\otimes {{\mathbf{I}}_{M}} \right)}^{*}}$, where $\mathbf{U}_{L}$ is the Fourier matrix with dimension $L$ given by 
\begin{equation}
\mathbf{U}_L=L^\frac{1}{2}\left[\exp{\left(\frac{-\text{i}2\pi\left(j-1\right)\left(k-1\right)}{L}\right)};\; j,k=1,\cdots,L\right],
\label{UL}
\end{equation} $\mathbf{I}_{M}$ is the identity matrix of order $M$ and  $\widehat{\mathbf{F}}=diag(\hat{\mathbf{F}}_1,\cdots, \hat{\mathbf{F}}_L)$ is a block diagonal matrix with each block $\hat{\mathbf{F}}_k$, $k=1, \cdots, L$ 
being an $M \times M$ matrix that contains all the spectral information associated to the frequency $w_k=\frac{k-1}{L}$.

\item Further diagonalization within  blocks. 
Each block is characterized by $M$ eigenvectors and $M$ eigenvalues.  
 
\begin{equation}
\hat{\mathbf{F}}_k=\mathbf{E}_k\hat{\mathbf{D}}_k\mathbf{E}_k^\prime
\label{F_k}
\end{equation}
with $\mathbf{E}_{k}^{{}}\mathbf{E}_{k}^{*}=\mathbf{E}_{k}^{*}\mathbf{E}_{k}^{{}}={{\mathbf{I}}_{M}}$. 

If we denote $\mathbf{E}=diag(\mathbf{E}_1\cdots \mathbf{E}_L)$, then the double diagonalization of $\mathbf{S}_\mathbf{C}$ gives as a result $\mathbf{S}_\mathbf{C}=\mathbf{V}\hat{\mathbf{D}}\mathbf{V}^*$, where 
\begin{equation}
\mathbf{V}=\left( {{\mathbf{U}}_{L}}\otimes {{\mathbf{I}}_{M}} \right)\mathbf{E}\in {{\mathbb{C}}^{LM\times LM}}
\label{V}
\end{equation}
and the block diagonal matrix $\mathbf{\hat{D}}=\operatorname{diag}\left( {{{\mathbf{\hat{D}}}}_{1}},\cdots ,{{{\mathbf{\hat{D}}}}_{L}} \right)$. Each block ${{\mathbf{\hat{D}}}_{k}}=\operatorname{diag}\left( {{{\hat{\lambda }}}_{k,1}},\cdots ,{{{\hat{\lambda }}}_{k,M}} \right)$ contains the eigenvalues of the  $\hat{\mathbf{F}}_k$ matrix in increasing order, ${{\hat{\lambda }}_{k,1}}\ge \cdots \ge {{\hat{\lambda }}_{k,M}}\ge 0$.
The notation of the columns (eigenvectors) of $\mathbf{V}$ as $\mathbf{v}_{j}=\mathbf{v}_{(k-1)M+m}=\mathbf{v}_{k,m}$ for $j=1,\cdots LM$, $k=1,\cdots, L$ and $m=1\cdots M$, emphasizes that the {\it j-th} eigenvector corresponds to the {\it k-th} frequency and the  {\it m-th} subcomponent (principal component within the block). That is, for each frequency $\omega_k$ there are $M$ eigenvectors.
\end{enumerate}
Notice that the Fourier unitary matrix defined in (\ref{UL}) has complex values. Therefore, we would like to make a change of basis so that the new set of eigenvectors that diagonalize $\mathbf{S}_C$, are all real. Proposition 2 (see Section 2.4.2), allows to substitute $\mathbf{V}$ by a real orthonormal basis $\mathbf{\tilde{V}}$ that diagonalizes $ \mathbf{S}_C$ with elements $\widetilde{\mathbf{V}}=\left[{\widetilde{\mathbf{v}}}_1|\cdots|{\widetilde{\mathbf{v}}}_{LM}\right]$, being ${\widetilde{\mathbf{v}}}_j={\widetilde{\mathbf{v}}}_{k,m}$ with $j=\left(k-1\right)M+m\quad, \forall \: k=1,\cdots,L$ and $m=1,\cdots,M$ defined by 
\begin{equation}
{{\mathbf{\tilde{v}}}_{k,m}}=\left\{ \begin{array}{*{35}{l}}
{{\mathbf{v}}_{k,m}} & k=1\text{ and }\tfrac{L}{2}+1\text{ if }L \text{ is even} \\
\sqrt{2}{{\mathcal{R}}_{{{\mathbf{v}}_{k,m}}}} & k=2,\cdots ,\left\lfloor \tfrac{L+1}{2} \right\rfloor \\
\sqrt{2}{{\mathcal{R}}_{{{\mathbf{v}}_{L+2-k,m}}}} & k=\left\lfloor \tfrac{L+1}{2}\right\rfloor +1,\cdots ,L\\
\end{array} \right.
\label{base_orto}
\end{equation}
where $\mathcal{R}_{\mathbf{v}}$ denotes the real part of the vector $\mathbf{v}$. 

Therefore, we can compute the elementary matrix for the {\it m-th} subcomponent at frequency $\omega_k$ as 
\begin{equation}
\mathbf{X}_{k,m}={\widetilde{\mathbf{v}}}_{k,m}{\widetilde{\mathbf{v}}}_{k,m}'\mathbf{X}\in\mathbb{R}^{LM\times N}.
\label{X_km}
\end{equation}
Consequently, the trajectory matrix can be decomposed as 
\begin{equation}
\mathbf{X}=\sum_{k,m}\mathbf{X}_{k,m}.
\label{Xsumkm}
\end{equation}
and the contribution to the total variability of the elementary matrix $\mathbf{X}_{k,m}$ is the ratio 
\begin{equation}
    \frac{{\hat{\lambda}}_{k,m}}{\sum_{k,m}{\hat{\lambda}}_{k,m}}.
    \label{contribution}
\end{equation} 

The piece of ${{\mathbf{\tilde{v}}}_{k,m}}$ corresponding to the series $(i)$ is denoted by $\mathbf{\tilde{v}}_{k,m}^{\left( i \right)}$ and is given by
\begin{equation}
    \mathbf{\tilde{v}}_{k,m}^{\left( i \right)}=\left( {{\mathbf{I}}_{L}}\otimes \mathbf{1}_{M,i}^\prime \right){{\mathbf{\tilde{v}}}_{k,m}}
    \label{v_piece}
\end{equation}
where ${{\mathbf{1}}_{M,i}}$ is a vertical vector of length $M$ with a $1$ in the {\it i-th} position and $0$ elsewhere. Therefore, the elementary matrix of the series $(i)$ for the {\it m-th} subcomponent of the frequency $\omega_k$ is obtained as
\begin{equation}
    \mathbf{X}_{k,m}^{\left( i \right)}=\mathbf{\tilde{v}}_{k,m}^{\left( i \right)}\mathbf{\tilde{v}}_{k,m}^\prime\mathbf{X}\in {{\mathbb{R}}^{L\times N}}.
    \label{Xi_km}
\end{equation}
In addition, the participation index of the series $(i)$ within the {\it m-th} subcomponent of the frequency $\omega_k$ is calculated, from (\ref{v_piece}), by the following expression
\begin{equation}
    \pi _{k,m}^{\left( i \right)}=\hat{\lambda }_{k,m}^{{}}{{\left( \mathbf{\tilde{v}}_{k,m}^{\left( i \right)} \right)}^{\prime}}\mathbf{\tilde{v}}_{k,m}^{\left( i \right)}.
    \label{participation}
\end{equation}

\bigskip
\textbf{3rd step: Grouping.}

The spectral density function of the vector process $\mathbf{x}_t$ is symmetric and, therefore, $\mathbf{\hat{F}}_{k}^{{}}=\mathbf{\hat{F}}_{L+2-k}^{\prime}$ for $k=2,\cdots ,\left\lfloor \tfrac{L+1}{2} \right\rfloor$. Then $\hat{\mathbf{D}}_k=\hat{\mathbf{D}}_{L+2-k}$ and, consequently, the subcomponents
$\mathbf{w}_{k,m}=\mathbf{X}'\widetilde{\mathbf{v}}_{k,m}$ and 
$\mathbf{w}_{L+2-k,m}=\mathbf{X}'\widetilde{\mathbf{v}}_{L+2-k,m}$ are harmonics of the same frequency. This motivates the creation of elementary pairs per subcomponent and frequency $B_{k,m}=\left\{\left(k,m\right),\left(L+2-k,m\right)\right\}$ for $k=2,\cdots,\left\lfloor \tfrac{L+1}{2} \right\rfloor$  except $B_{1,m}=\left\{\left(1,m\right)\right\}$ and occasionally $B_{\frac{L}{2}+1,m}=\left\{\left(\frac{L}{2}+1,m\right)\right\}$ if $L$ is even. The matrices corresponding to the pairs $B_{k,m}$ are given by the sum of two elementary matrices per subcomponent and frequency
\begin{equation}
\mathbf{X}_{B_{k,m}}=\mathbf{X}_{k,m}+\mathbf{X}_{L+2-k,m}.
\label{X_Bkm}
\end{equation}

Therefore, both the matrices associated with the elementary pairs $B_{k,m}$ by subcomponent and frequency, and the oscillatory components obtained from them are previously identified with a determined frequency as in the univariate case. 

Under the assumption of separability, we define $G$ disjoint groups of the elementary pairs per subcomponent and frequency. The resulting matrix for each of the disjoint groups is defined as the sum of the matrices associated with the pairs $B_{k,m}$ included. If $I_j=\left\{ B_{k_{j_1},m_{j_1}}, ..., B_{k_{j_q},m_{j_q}}\right\}$, $j=1,...,G$ is each disjoint group of $j_q$ pairs $B_{k,m}$ with $ 1\le j_q\le \left\lfloor \tfrac{L+1}{2} \right\rfloor M$, then the matrix 
$\mathbf{X}_{I_j}$ from group $I_j$ is calculated as the sum of the corresponding matrices defined by (\ref{X_Bkm}), 
$ \mathbf{X}_{I_j}=\sum_{B_{k,m}\in I_j}\mathbf{X}_{B_{k,m}}$ and the trajectory matrix $\mathbf{X}$ can be decomposed as the sum of these group matrices
\begin{equation*}
\mathbf{X}=\mathbf{X}_{I_1}+\mathbf{X}_{I_2}+\cdots+\mathbf{X}_{I_G}.
\label{3_31} 
\end{equation*}

If we wish to extract the oscillation at frequency $\frac{k-1}{L}$, the appropriate group is $B_k=\left\{B_{k,m}\; \forall\: m, 1\le m\le M\right\} $ being the resulting matrix
$ \mathbf{X}_{B_k}=\sum_{m=1}^{M}\mathbf{X}_{B_{k,m}}$

\bigskip
\textbf{4th step: Reconstruction.}

Finally, each $L\times\ N$ matrix ${{\mathbf{X}}_{{{I}_{j}}}}$ of vectors $M\times 1$ from the previous step is transformed into a new time series of length $T$ by diagonal averaging, producing the reconstructed time series $\widetilde{\mathbf{x}}_{I_j,t}=\left({\widetilde{x}}_{I_j,t}^{\left(1\right)},\cdots,\widetilde{x}_{I_j,t}^{\left(M\right)}\right)'$. If $\mathbf{x}_{r,s}^{I_j}$ are the vector elements of the matrix $\mathbf{X}_{I_j}$, then the values of the reconstructed vector time series 
$ \widetilde{\mathbf{x}}_{I_j,t}$, with $L<N$ are calculated as in Vautard et al. (1992) but the formula is adapted to vectors to ``hankelize'' the matrix 
$\mathbf{X}_{I_j}$ with the operator $\text{H}(\centerdot)$ as follows:

\begin{equation}
{{\mathbf{\tilde{x}}}_{{{I}_{j}},t}}=\text{H}\left(\mathbf{X}_{I_j}\right) = \left\{
\begin{array}{*{35}{l}}
\tfrac{1}{t}\sum\limits_{i=1}^{t}{\mathbf{x}_{i,t-i+1}^{{{I}_{j}}}} & 1\le t<L \\
\tfrac{1}{L}\sum\limits_{i=1}^{L}{\mathbf{x}_{i,t-i+1}^{{{I}_{j}}}} & L\le t\le N \\
\tfrac{1}{T-t+1}\sum\limits_{i=t-T+L}^{L}{\mathbf{x}_{i,t-i+1}^{{{I}_{j}}}} & N<t\le T \\
\end{array} \right.
\label{hankelize}
\end{equation}

The reconstructed multivariate times series resulting from $B_{k,m}$ are called elementary reconstructed vector time series by subcomponent $m$ and frequency $k$.


The pseudo-code with the logical sequence of the details of the four steps of M-CiSSA is provided in Algorithm 1.

\spacingset{1}
\begin{algorithm}
\label{Code_CiSSA}
\caption{Pseudo-code of Multivariate Circulant SSA}
\begin{algorithmic}[1]
\Require Time series multivariate ${{\mathbf{x}}_{t}}$ and window length $L$
\Ensure Reconstructed subcomponents associated with each disjoint group of frequencies
\State Construct the trajectory matrix $\mathbf{X}$ by (\ref{eq_trajectory})
\For{$k=0$ \textbf{to} $L-1$} \Comment \textit{Estimated autocovariance matrices}
	\State Compute the lagged autocovariance matrices ${{\mathbf{\hat{\Gamma}}}_{k}}$
\EndFor
\For{$k=0$ \textbf{to} $L-1$} \Comment \textit{First row of block circulant matrix $\mathbf{S}_{C}$}
	\State Compute ${{\mathbf{\hat{\Omega }}}_{k}}$ given in (\ref{omega_tilde})
\EndFor
\State Build the block circulant matrix $\mathbf{S}_{C}$
\For{$k=1$ \textbf{to} $L$} \Comment \textit{Double diagonalization of $\mathbf{S}_{C}$ and elementary matrices}
	\State Find the eigenvalues $\widehat{\lambda}_{k,m}$ of $\mathbf{S}_\mathbf{C}={\mathbf{V}}\hat{\mathbf{D}}{{\mathbf{V}}}^*$ where $\mathbf{D}_k=\operatorname{diag}{\left(\widehat{\lambda}_{k,1},\cdots,\widehat{\lambda}_{k,M}\right)}$.
	\State Calculate its corresponding eigenvectors ${\mathbf{\tilde{v}}}_{k,m}$ following (\ref{UL}), (\ref{V}) and (\ref{base_orto})
	\State The pair $(\widehat{\lambda }_{k,m}, {\mathbf{\tilde{v}}}_{k,m})$ is attached to the frequency $w_k=\frac{k-1}{L}$ and subcomponent $m$
	\State Compute the elementary matrix $\mathbf{X}_{k,m}={\widetilde{\mathbf{v}}}_{k,m}{\widetilde{\mathbf{v}}}_{k,m}'\mathbf{X}$ associated with $(w_k,m)$
        \State Determine the contribution of the elementary matrix $\mathbf{X}_{k,m}$ by (\ref{contribution})
\EndFor
\State Set the group $B_{1,m}=\{(1,m)\}$ and the matrix $\mathbf{X}_{B_{1,m}}=\mathbf{X}_{1,m}$ \Comment \textit{Elementary pairs and matrices by frequency and subcomponent}
\For{$k=2$ \textbf{to} $\left\lfloor \tfrac{L+1}{2} \right\rfloor$} 
	\State Set elementary pair by frequency and subcomponent $B_{k,m}=\{(k,m),(L+2-k,m)\}$
	\State Compute elementary matrix by frequency and subcomponent  $\mathbf{X}_{B_{k,m}}$ by (\ref{X_Bkm})
\EndFor
\If{$L$ is even}
	\State Set the group $B_{L/2+1,m}=\left\{ (L/2+1,m) \right\}$ and the matrix $\mathbf{X}_{B_{L/2+1,m}}=\mathbf{X}_{L/2+1,m}$
\EndIf
\State Determine the $G$ disjoint groups $I_j$ of the elementary pairs $B_{k,m}$ with the interesting frequencies $w_k$ and subcomponents $m$ for the non-zero contributions
\For{$j=1$ \textbf{to} $G$} \Comment \textit{Matrices associated with the disjoint groups}
	\State Compute the matrix $\mathbf{X}_{I_{j}}$ associated with each group $I_j$ by $\mathbf{X}_{I_{j}}=\sum_{B_{k,m}\in I_j} \mathbf{X}_{B_{k,m}}$
\EndFor
\For{$j=1$ \textbf{to} $G$} \Comment \textit{Reconstructed series}
	\State Calculate the reconstructed series ${{\mathbf{\tilde{x}}}_{{{I}_{j}},t}}$ with the \textit{hankelization} of matrix $\mathbf{X}_{I_j}$ by (\ref{hankelize})
\EndFor
\end{algorithmic}
\end{algorithm}
\spacingset{1.45}

The M-CiSSA algorithm described so far requires stationary time series. However, it is straightforward to show that it can also be applied to non-stationary time series. Other versions of multivariate SSA, Basic SSA (Broomhead and King, 1986b) and Toeplitz SSA (Plaut and Vautard, 1994), are also implemented on non-stationary time series. 
In the case of Circulant SSA, Bógalo et al. (2021) prove its validity for non-stationary univariate time series approximating the discontinuities of the spectrum by means of a pseudo-spectrum. 
The same approximation can be applied in the case of $M$ series instead of just one. 
More recently, there are examples of applying alternative versions of SSA to non-stationary multivariate time series (see, e.g., Groth et al., 2011; Carvalho (de) and Rua, 2017). In a related but different context, Peña and Yohai (2016) also use principal components and lagged principal components with non-stationary time series.

\subsection{Pillars of M-CiSSA}
\subsubsection{Building the block circulant matrix}
Consider the sequence of lagged cross variance-covariance matrices of the population as a function of the window length $L$ and denote it by $\mathbf{T}_L$. Each matrix in that sequence is an $L\times L$ block Toeplitz matrix with $M\times M$ blocks resulting in a $LM\times LM$ Hermitian matrix. Let, 
\begin{equation}
\mathbf{T}_L=\left[\mathbf{\Gamma}_{ij}=\mathbf{\Gamma}_{i-j};i,j=1,\cdots,L\right].
\label{T_L}
\end{equation}
It is well known that the sequence $\left\{ \mathbf{\Gamma}_k \right\} _{k\in\mathbb{Z}}$ can be generated as
\begin{equation*} 
\mathbf{\Gamma}_k=\int_{0}^{1}\mathbf{F}\left(\omega\right)\exp{\left(-\text{i}2\pi k\omega\right)}d\omega,\;\forall k\in\mathbb{Z}
\end{equation*} 
where $\omega\in\left[0,1\right]$ is the frequency in cycles per unit of time, $\text{i}=\sqrt{-1}$ is the imaginary unit and $\mathbf{F}\left(\omega\right)$ is the spectral density matrix of the stochastic process of the vector time series $\mathbf{x}_t$, that is, the ''matrix'' Fourier series given by
\begin{equation}
\mathbf{F}\left(\omega\right)=\sum_{k=-\infty}^{\infty}{\mathbf{\Gamma}_k\exp{\left(\text{i}2\pi k\omega\right)}},\; \omega\in\left[0,1\right].
\label{F_w}
\end{equation}
The sequence $\left\{\mathbf{\Gamma}_k\right\}_{k\in\mathbb{Z}}$ are the Fourier coefficients of the matrix-valued function $\mathbf{F}\left(\omega\right)$. As a consequence, the continuous and $2\pi$-periodic matrix-valued function $\mathbf{F}\left(\omega\right)$ of a real variable is the generating function or symbol of the matrix $\mathbf{T}_L\left(\mathbf{F}\right)=\mathbf{T}_L$ that originates the sequence of block Toeplitz matrices that we denote $\left\{\mathbf{T}_L\left(\mathbf{F}\right)\right\}$.

With this approach, we do not have a closed formula for the eigenvalues and eigenvectors of the Toeplitz matrix of second moments given in (\ref{T_L}). This problem can be solved, if instead of block Toeplitz matrices, we use block circulant matrices.

Let $\mathbf{C}_L$ be an $L\times L$ block circulant matrix with blocks $M\times M$, that is, $\mathbf{C}_L$ is an $LM\times LM$ matrix of the form

\begin{equation}
\mathbf{C}_L=\left(\begin{matrix}\mathbf{\Omega}_0&\mathbf{\Omega}_1&\mathbf{\Omega}_2&\cdots&\mathbf{\Omega}_{L-1}\\\mathbf{\Omega}_{L-1}&\mathbf{\Omega}_0&\mathbf{\Omega}_1&\ddots&\vdots\\\mathbf{\Omega}_{L-2}&\mathbf{\Omega}_{L-1}&\mathbf{\Omega}_0&\ddots&\mathbf{\Omega}_2\\\vdots&\ddots&\ddots&\ddots&\mathbf{\Omega}_1\\\mathbf{\Omega}_1&\cdots&\mathbf{\Omega}_{L-2}&\mathbf{\Omega}_{L-1}&\mathbf{\Omega}_0\\\end{matrix}\right)
 \label{C_L}
\end{equation}
where $\mathbf{\Omega}_k\in\mathbb{C}^{M\times M},\; k=0,1,\cdots,L-1$. We say that $\mathbf{C}_L$ is block circulant since each block row is built from a right shift of the blocks of the previous block row.
Each block of $\mathbf{C}_L$ can be generated according to Gutiérrez-Gutiérrez and Crespo (2008) as 
\begin{equation*}
\mathbf{\Omega}_k=\frac{1}{L}\sum_{j=0}^{L-1}{\mathbf{F}\left(\frac{j}{L}\right)\exp{\left(\frac{\text{i}2\pi jk}{L}\right)}},\; k=0,\cdots,L-1
\label{omega}
\end{equation*}
being $\mathbf{F}\left(\omega\right)$ the generating function of the sequence $\left\{\mathbf{T}_L\left(\mathbf{F}\right)\right\}$ that also generates the sequence of block circulant matrices $\left\{\mathbf{C}_L\left(\mathbf{F}\right)\right\}$.
Moreover, as Gutiérrez-Gutiérrez and Crespo (2008), Lemma 6.1, show the two block matrices sequences $\left\{\mathbf{T}_L\left(\mathbf{F}\right)\right\}$ and $\left\{\mathbf{C}_L\left(\mathbf{F}\right)\right\}$ are asymptotically equivalent as $L\rightarrow \infty $, $\mathbf{T}_{L}(\mathbf{F})$ $\sim $ $\mathbf{C}_{L}(\mathbf{F})$, 
in the sense that both matrices have
bounded eigenvalues and $\underset{L\rightarrow \infty }{\lim }\frac{
\left\Vert \mathbf{T}_{L}(\mathbf{F})-\mathbf{C}_{L}(\mathbf{F})\right\Vert_{F}}{\sqrt{L
}}=0$, where $\left\Vert \text{\textperiodcentered }\right\Vert_{F}$ is the Frobenius norm.

The advantage of using the block circulant matrix $\mathbf{C}_L$ instead of the block Toeplitz matrix $\mathbf{T}_{L}$ is that the former can be block diagonalized, while the later cannot. However, in order to build the block matrix (\ref{C_L}), we should either know the matrix function $\mathbf{F}$ or the infinite sequence $\left\{\mathbf{\Gamma}_k\right\}_{k\in\mathbb{Z}}$. We realize that, in practice, we will have a finite number of second order matrices so, in order to make this approach operational, we generalize the results given by Pearl (1973) for the continuous and $2\pi$-periodic scalar function $f$ to the continuous and $2\pi$-periodic matrix function $\mathbf{F}$. In particular, similarly to the proposal put forward by Pearl (1973) for $1 \times 1 $ blocks, we suggest using as $M \times M$ blocks of the first row in $\mathbf{C}_L$ given by (\ref{omega_tilde}):
\begin{equation*}
{\widetilde{\mathbf{\Omega}}}_k=\frac{k}{L}\mathbf{\Gamma}_{L-k}+\frac{L-k}{L}\mathbf{\Gamma}_{-k}, \; k=0,\cdots,L-1.
\end{equation*}
This way of defining the block circulant matrix $\mathbf{C}_L$ is associated with the continuous and $2\pi$-periodic matrix function ${\widetilde{\mathbf{F}}}$,
\begin{equation}
{\widetilde{\mathbf{F}}}\left(\omega\right)=\frac{1}{L}\sum_{m=1}^{L}\sum_{l=1}^{L}{\mathbf{\Gamma}_{l-m}\exp{\left(\text{i}2\pi\left(l-m\right)\omega\right)}},\;\omega\in\left[0,1\right],
\label{F_tilde}
\end{equation}
that generates the sequence of block circulant matrices $\left\{\mathbf{C}_L\left(\widetilde{\mathbf{F}}\right)\right\}$. Theorem \ref{Theorem_1} shows the asymptotic equivalence between the sequences $\left\{\mathbf{T}_L\left(\mathbf{F}\right)\right\}$ and $\left\{\mathbf{C}_L\left(\widetilde{\mathbf{F}}\right)\right\}$, denoted as
$\mathbf{T}_{L}(\mathbf{F})$ $\sim $ $\mathbf{C}_{L}(\widetilde{\mathbf{F}})$. This theorem is the basis of our proposed algorithm since we will use the later sequences in our Multivariate Circulant SSA proposal, which we have called M-CiSSA.

\begin{theorem} \label{Theorem_1}
Let $\mathbf{F}:\left[0,1\right]\rightarrow\mathbb{C}^{M\times N}$ be a matrix-valued function of real variable which is continuous and 2$\pi$-periodic and let ${\widetilde{\mathbf{F}}}$ be the matrix-valued function defined in (\ref{F_tilde}) from the Fourier coefficients of the matrix-valued function $\mathbf{F}$, then, $\mathbf{T}_L\left(\mathbf{F}\right)\sim\mathbf{C}_L\left({\widetilde{\mathbf{F}}}\right)$.
\end{theorem}

\begin{proof}
The proof is given in the appendix.
\end{proof}

\bigskip

\subsubsection{Double diagonalization of the block circulant matrix}

Once the block circulant matrix is defined, we check the properties of its diagonalization. 
In particular, we see that the diagonal blocks are associated with known frequencies and that further diagonalizing within the blocks is possible. 
As a consequence, these results allow us to understand fluctuations by frequency and within frequency.

To do so, we start by showing how the block circulant matrix sequence $\left\{\mathbf{C}_L\left(\mathbf{F}\right)\right\}$ is diagonalized.
For any matrix valued function $\mathbf{F}:\left[0,1\right]\rightarrow\mathbb{C}^{M\times M}$ which is continuous and $2\pi$-periodic, the block circulant matrix $\mathbf{C}_L\left(\mathbf{F}\right)$ is characterized according to Gutiérrez-Gutiérrez and Crespo (2008) by a block diagonalization given by
\begin{equation*}
\mathbf{C}_L\left(\mathbf{F}\right)=\left(\mathbf{U}_L\otimes\mathbf{I}_M\right)\operatorname{diag}{\left(\mathbf{F}_1,\cdots,\mathbf{F}_L\right)}\left(\mathbf{U}_L\otimes\mathbf{I}_M\right)^*
\label{diag_C_L}
\end{equation*}
where $\mathbf{U}_L$ is given by (\ref{UL}). Each block $\mathbf{F}_k=\mathbf{F}\left(\frac{k-1}{L}\right)$, $k=1,\cdots,L$ represents the cross spectral density matrix of the multivariate stochastic process $\mathbf{x}_t$ for the frequency $\omega_k=\frac{k-1}{L}, k=1,\cdots,L$ and can be unitarily diagonalized. In this way, we obtain  $\mathbf{F}_k=\mathbf{E}_k\mathbf{D}_k\mathbf{E}_k^*$ with $\mathbf{E}_k\in\mathbb{C}^{M\times M}$ and $\mathbf{E}_k\mathbf{E}_k^*=\mathbf{E}_k^*\mathbf{E}_k=\mathbf{I}_M$, where $\mathbf{E}_k=\left[\mathbf{e}_{k,1}|\cdots|\mathbf{e}_{k,M}\right]$ contains the eigenvectors, and the diagonal matrix $\mathbf{D}_k=\operatorname{diag}{\left(\lambda_{k,1},\cdots,\lambda_{k,M}\right)}$ contains the ordered eigenvalues $\lambda_{k,1}\geq\cdots\geq\lambda_{k,M}\geq0$ of $\mathbf{F}_k$. Therefore, the unitary diagonalization of the Hermitian matrix $\mathbf{C}_L\left(\mathbf{F}\right)$ is given by $\mathbf{C}_L\left(\mathbf{F}\right)=\mathbf{VD}\mathbf{V}^*$ with 
\begin{equation*}
\mathbf{V}=\left(\mathbf{U}_L\otimes\mathbf{I}_M\right)\mathbf{E}\in\mathbb{C}^{LM\times L M} \label{matrix_V}
\end{equation*}
where $\mathbf{E}=\operatorname{diag}{\left(\mathbf{E}_1,\cdots,\mathbf{E}_L\right)}$ and $\mathbf{D}=\operatorname{diag}{\left(\mathbf{D}_1,\cdots,\mathbf{D}_L\right)}$. As a consequence, there are $M$ eigenvectors associated with each frequency $\omega_k=\frac{k-1}{L}$. The {\it j-th} eigenvector $\mathbf{v}_j,\; j=1,...,LM$ of the matrix $\mathbf{C}_L\left(\mathbf{F}\right)$ is given by 
\begin{equation*}
\mathbf{v}_j=\mathbf{v}_{\left(k-1\right)M+m}=\mathbf{v}_{k,m}=\mathbf{u}_k\otimes\mathbf{e}_{k,m}
\end{equation*}
for $k=1,\cdots,L$ and $m=1,\cdots, M$ where $\mathbf{u}_k$ is the {\it k-th} column of the Fourier unitary matrix $\mathbf{U}_L$ and $\mathbf{e}_{k,m}$ is the {\it m-th} eigenvector of the cross spectral density matrix $\mathbf{F}_k$. Notice that $ \mathbf{F}$ is symmetric with respect to the frequency $ \frac{1}{2}$ as deduced from (\ref{F_w}). 
This means $\mathbf{F}_k=\mathbf{F}_{L+2-k}^{\prime};\; k=2,\cdots,\left\lfloor\frac{L+1}{2}\right\rfloor$. So the corresponding eigenvectors are conjugated $ \mathbf{E}_k=\overline{\mathbf{E}}_{L+2-k}$ and the associated eigenvalues are equal $ \mathbf{D}_k=\mathbf{D}_{L+2-k}$.
Proposition \ref{Proposition_1} states how to orthogonally diagonalize the circulant matrix $\mathbf{C}_L\left(\mathbf{F}\right)$.

\begin{proposition} \label{Proposition_1}
Let $\mathbf{C}_L\left(\mathbf{F}\right)$ be the block circulant matrix given by (\ref{diag_C_L}) and let $ \mathbf{V}$ be the unitarily matrix obtained by 
\begin{equation*}
{{\mathbf{\tilde{v}}}_{k,m}}=\left\{ \begin{array}{*{35}{l}}
{{\mathbf{v}}_{k,m}} & k=1\text{ and }\tfrac{L}{2}+1\text{ if }L \text{ is even} \\
\sqrt{2}{{\mathcal{R}}_{{{\mathbf{v}}_{k,m}}}} & k=2,\cdots ,\left\lfloor \tfrac{L+1}{2} \right\rfloor \\
\sqrt{2}{{\mathcal{R}}_{{{\mathbf{v}}_{L+2-k,m}}}} & k=\left\lfloor \tfrac{L+1}{2}\right\rfloor +1,\cdots ,L\\
\end{array} \right.
\end{equation*}
that unitary diagonalizes $ \mathbf{C}_L\left(\mathbf{F}\right)$. The set of vectors $ \left\{{\widetilde{\mathbf{v}}}_{k,m}\right\}_{k,m=1}^{L,M}$ defined $\forall\,m=1,\cdots, M $ by (\ref{base_orto})
is an orthonormal basis of $ \mathbb{R}^{LM} $ that orthogonally diagonalizes the matrix $ \mathbf{C}_L\left(\mathbf{F}\right)$, $\mathbf{C}_L\left(\mathbf{F}\right)=\widetilde{\mathbf{V}}\mathbf{D}\widetilde{\mathbf{V}}'$, where  $\widetilde{\mathbf{V}}=\left[{\widetilde{\mathbf{v}}}_1|\cdots|{\widetilde{\mathbf{v}}}_{LM}\right]$, being ${\widetilde{\mathbf{v}}}_j={\widetilde{\mathbf{v}}}_{k,m}$ with $j=\left(k-1\right)M+m\quad \forall \: k=1,\cdots,L$ and $m=1,\cdots,M$.
\end{proposition}

\begin{proof}
The proof is given in the appendix.
\end{proof}

\subsection{Understanding the formation of univariate signals in the multivariate setup: uniqueness between CiSSA and M-CiSSA}

Given the {\it i-th} variable of the set of time series, its corresponding signal for a particular frequency can be estimated in two different ways: in a univariate framework (or M-CiSSA with $M=1$) or within the multivariate setup. 
In this section, we prove that the univariate signal is the result of adding up the $M$ subcomponents within a given frequency for that series. This is relevant, for instance, to understand the formation of the cycles associated to each time series.
While we could extract the cycle within the univariate framework, in doing so within the multivariate setup allows us to decompose the univariate cycle as sum of subcomponents that reflect the relation among the different variables. 
In particular, we are able to disentangle which part of the univariate cycle is common to the other variables and which part is specific or idyosincratic and not shared with the rest of the variables.

The univariate estimation in CiSSA of the oscillatory component of the {\it i-th} series for each frequency ${{\omega }_{k}}=\tfrac{k-1}{L}$, $k=1,\cdots ,L$, 
originates a single time series or component associated with the elementary matrix by frequency $\mathbf{X}_{k}^{\left( i \right)}$. However, the estimation of that same oscillatory component by M-CiSSA produces $M$ series or subcomponents respectively associated with the $ M$ elementary matrices by subcomponent and frequency $\mathbf{X}_{k,m}^{\left( i \right)}$. The sum of these matrices, $\sum\limits_{m=1}^{M}{\mathbf{X}_{k,m}^{\left( i \right)}}$, originates the estimation in M-CiSSA of the oscillatory component of the {\it i-th} series at frequency ${{\omega }_{k}}$. Theorem \ref{Theorem_2} proves that the two signals are the same.

\begin{theorem}[Uniqueness]\label{Theorem_2}
Given a window length $L$, the oscillatory components derived from the matrices $\mathbf{X}_{k}^{\left( i \right)}$ and $\sum\limits_{m=1}^{M}{\mathbf{X}_{k,m}^{\left( i \right)}}$
of any {\it i-th} time series at each frequency ${{\omega }_{k}}$ obtained with univariate and multivariate CiSSA, respectively, are identical. That is, 
the oscillatory component of any {\it i-th} time series for each frequency ${{\omega }_{k}}$ is unique.
\end{theorem}

\begin{proof}
The proof is given in the appendix.
\end{proof}

\bigskip
This result allows us to use M-CiSSA as a way to de-noise the extracted signals and also to obtain the common spectral signals and estimate co-movements.
Regarding de-noising, M-CiSSA allows us to separate the signal from an over imposed colored noise, as defined by Allen and Robertson (1996), by selecting a reduced number of components associated with the non-null eigenvalues estimated for the frequency $\omega_k$. In general, to estimate the signal of a harmonic, it will be enough to select a small number of subcomponents associated with their higher eigenvalues that will characterize both the amplitude and the dating of the oscillatory components. In this sense, the subcomponents help to extract the common spectral signals as well as the co-movements in order to analyze their characterization (procyclical or anticyclical) and cyclical position (leading, coincident, or lagging) of the oscillatory components as Groth et al (2011) observed. Therefore, these subcomponents describe the formation of the oscillatory components in a multivariate setup.

Theorem \ref{Theorem_2} also proves the empirical result by Plaut and Vautard (1994) that, in M-SSA, an oscillatory pair does not explain all the variability due to a harmonic. Classical versions of M-SSA only consider the highest eigenvalues and omit information related to that harmonic. This information may hold great interest for economic analysis because it allows us to observe, for each of the series, the different gaps for the same oscillatory component.

\section{Applications}
\subsection{A synthetic example}
\begin{figure}[ht]
\begin{center}
\includegraphics[width=5.7in]{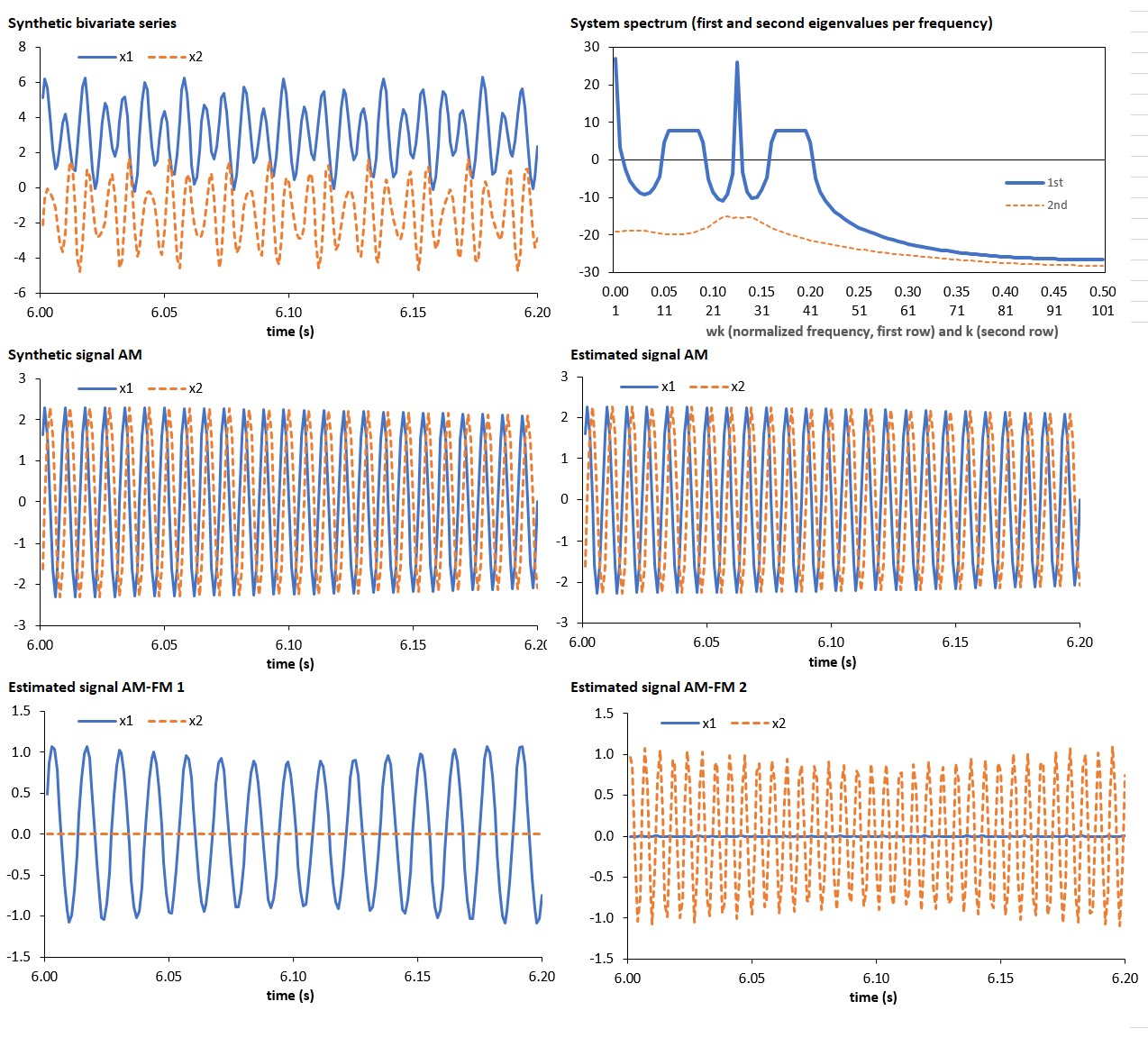}
\end{center}
\caption{Synthetic bivariate example: Simulated series, estimated power spectral density and estimated signals.} \label{sint_1}
\end{figure}

The following synthetic example shows the performance of M-CiSSA with various types of signals. We  consider two ($M=2$) signals $x_{1}(t)$ and $x_{2}(t)$ each one of them generated as the sum of a linear trend $T_{i}(t)$ plus a signal modulated in amplitude $S_{i}(t)$ plus another oscillatory component modulated both in amplitude and frequency $Y_{i}(t)$, such that $x_{i}(t)=T_{i}(t)+S_{i}(t)+Y_{i}(t)$, $i=1,2$. The trend has positive slope for the first series $T_{1}(t)=0.5t$ and negative for the second one $T_{2}(t)=-0.25t$.  The AM components are generated as $S_{1}(t)=A_{S}(t)sin(\omega_{S}t)$ and $S_{2}(t)=A_{S}(t)sin(\omega_{S}t-\pi/2)$ being out of phase $\pi/2$. Finally, the AM-FM component are generated as $Y_{i}(t)=A_{Y}(t)sin(\omega_{Y_{i,a}} t+ \omega_{Y_{i,b}}\frac{t^2}{2T})$, $i=1,2$. The modulated amplitudes are generated as $A_{S}(t)=2+0.3cos(\omega_{A,S} t)$ and $A_{Y}(t)=1+0.1cos(\omega_{A,Y} t)$. Notice that the frequency modulated signals show linearly increasing frequencies in time, $\omega_{Y_{i}}=\omega_{Y_{i,a}} + \omega_{Y_{i,b}}\frac{t}{2T}$, $i=1,2$. The chosen values for the frequencies are $f_{S}=125Hz$, $f_{A,S}=1Hz$, $f_{A,Y}=5Hz$, $f_{Y_{1,a}}=50Hz$, $f_{Y_{2,a}}=180Hz$ and $f_{Y_{1,b}}=f_{Y_{2,b}}=40Hz$, being $\omega=2\pi f$. Similar synthetic signals have been used, for example, in  Biagietti et al. (2015) and Bógalo et al. (2021).
The sampling frequency is 1000$Hz$ and the signals are observed for 10 seconds, therefore the number of observations is 10000. 
The left panels in Figure \ref{sint_1}, show the simulated time series (first row) and the AM and AM-FM signals (second and third rows respectively) for a span of 2 seconds.

We choose $L=200$  and perform the block eigendecomposition by frequency that contains the spectral information associated to the frequency $w_k=\frac{k-1}{L}, k=1, ..., 200$.
Each block is characterized by the 2 eigenvalues $\hat{\lambda}_{k,m}, m=1,2$ of the diagonal matrix $\hat{\textbf{D}}_k, k=1, ..., 200$ that also define the contribution to the total variability of the bivariate system as in (\ref{contribution}).
The top right panel of Figure 1 shows the estimation of the spectral density of the bivariate system by the first and second (dynamic) eigenvalues, measured in dB, for each frequency, where in the x-axis we have represented both, the values of $k$ and the equivalent normalized frequencies $w_k$ to highlight the automated identification that M-CiSSA provides.

First, it can be seen that in this particular example the first eigenvalue dominates the second one for every frequency. 
Therefore, it would be enough to just analyze the first eigenvalue, instead of the trace for each block, to identify the most relevant frequencies of fluctuation. Notice that all the values of the second eigenvalue are negative (the scale is logarithmic) and, therefore, its information content is negligible.
Second, it clearly shows a peak at the zero frequency $(k=1)$ that captures the linear trend and another one at the normalized frequency of 0.125 ($k=26$) that corresponds to $S_{i}(t), i=1,2$ and the 2 "plateau" corresponding to the frequency modulated signals $Y_1(t)$ and $Y_2(t)$. The first "plateau" goes from $k=11$ to $k=19$ and the second one from $k=33$ to $k=41$, that translated into frequencies correspond to the normalized frequencies between 0.05  to 0.07 for the first one and frequencies between 0.18 and 0.20 for the second one.
Therefore, M-CiSSA will capture the modulated frequency by adding components of adjacent frequencies.
These components are the more relevant in the bivariate system and they account for 66.25\% of the total variability.

Once the analysis of the block diagonalization is complete, we study the within blocks diagonalization.
Our goal in this second step  is to isolate and reconstruct the latent signals as well as to know how much each of the observed series or channels contributes to that reconstruction. 
To illustrate that M-CiSSA is accurate to reproduce AM signals, the middle row of Figure \ref{sint_1} shows the synthetic signals $S_1(t)$ and $S_2(t)$ on the left column and their reconstruction on the right one, for a period of 2 seconds. 
The two signals are reconstructed using, first, (\ref{X_km}) for $k=26$ and its symmetric value in $L+2-k=176$ which correspond to the frequency of 125Hz; and then, the antidiagonal averaging given in (\ref{hankelize}). The plot in the middle column, right panel shows the two columns of the reconstructed series ${{\mathbf{\tilde{x}}}_{{{I}_{j}},t}}$ for the same time span of 2 seconds. 
For the AM-FM signals the same procedure is applied for $k=11$ to $k=19$ and their symmetric counterparts (left, bottom panel) where we show the high fluctuations of the first plateau for the reconstruction of $Y_1(t)$, being negligible those linked to the reconstruction of $Y_2(t)$; and $k=33$ to $k=41$ (right, bottom panel) and, again, their symmetric counterparts, where the variation in $Y_2(t)$ is shown\footnote{The reconstruction of the trend follows the same methodology with $k=0$ and exactly reproduces the generated ones. Results are available from the authors upon request.}. 

 All in all, we have seen that M-CiSSA is able to capture latent components of very different nature (non-stationary, common, idiosyncratic, modulated in amplitude, with time varying frequency), that can be out of phase, reconstructing the sources of variation in a multivariate context. The introduced circulant matrices of second moments allows the match between frequencies and eigenvalues, providing an automated identification of the extracted signals.  

\subsection{Primary Commodity Energy Prices}

We now apply M-CiSSA to the Primary Commodity Energy Prices published by the International Monetary Fund (IMF). Energy commodities accounted for 40.9\% of the world trade between 2014 and 2016 and are central to competitiveness in industry, notably influencing consumers by affecting their energy consumption patterns and total expenditure. 
The evolution of energy prices impacts on other non-energy primary commodities (Kratschell
and Schmidt, 2017), exchange rates (Xu et al., 2019), and inflation (Garratt and Petrella, 2019), among others. 
Energy prices not only affect economic activity but they are also
influenced by it (as Alquist et al., 2019, and Kilian and Zhou, 2018, show for commodity prices in general). 
Also real price shifts may affect the commodity demand and result in one of the causes of fluctuations of the business cycle. 
This relationship suggests that besides analyzing the long-run behaviour
for policy issues, studying the cyclical frequency of energy prices is also of paramount importance.
A second issue is related to decoupling among different energy prices, as on
a theoretical and empirical basis, oil and natural gas are close substitutes in the long run. 
In this regard, some authors have found that oil prices drove US gas prices, and also that
oil prices led the co-movement between European and North American natural gas prices (see, e.g., Brown and Yucel, 2008, 2009). 
Despite this evidence, US oil and gas prices seem to have decoupled since 2009 triggering a new discussion on whether or not this is
permanent (see, e.g., Erdos, 2012; Nick and Thoenes, 2014; Zhang and Ji, 2018). Since
target policies should differ depending on whether decoupling amongst markets
occurs in the long run or at medium frequency, we have tackled this problem by addressing the
particular frequencies at which markets might be decoupled.
 
\begin{figure}[ht]
\begin{center}
\includegraphics[width=5in]{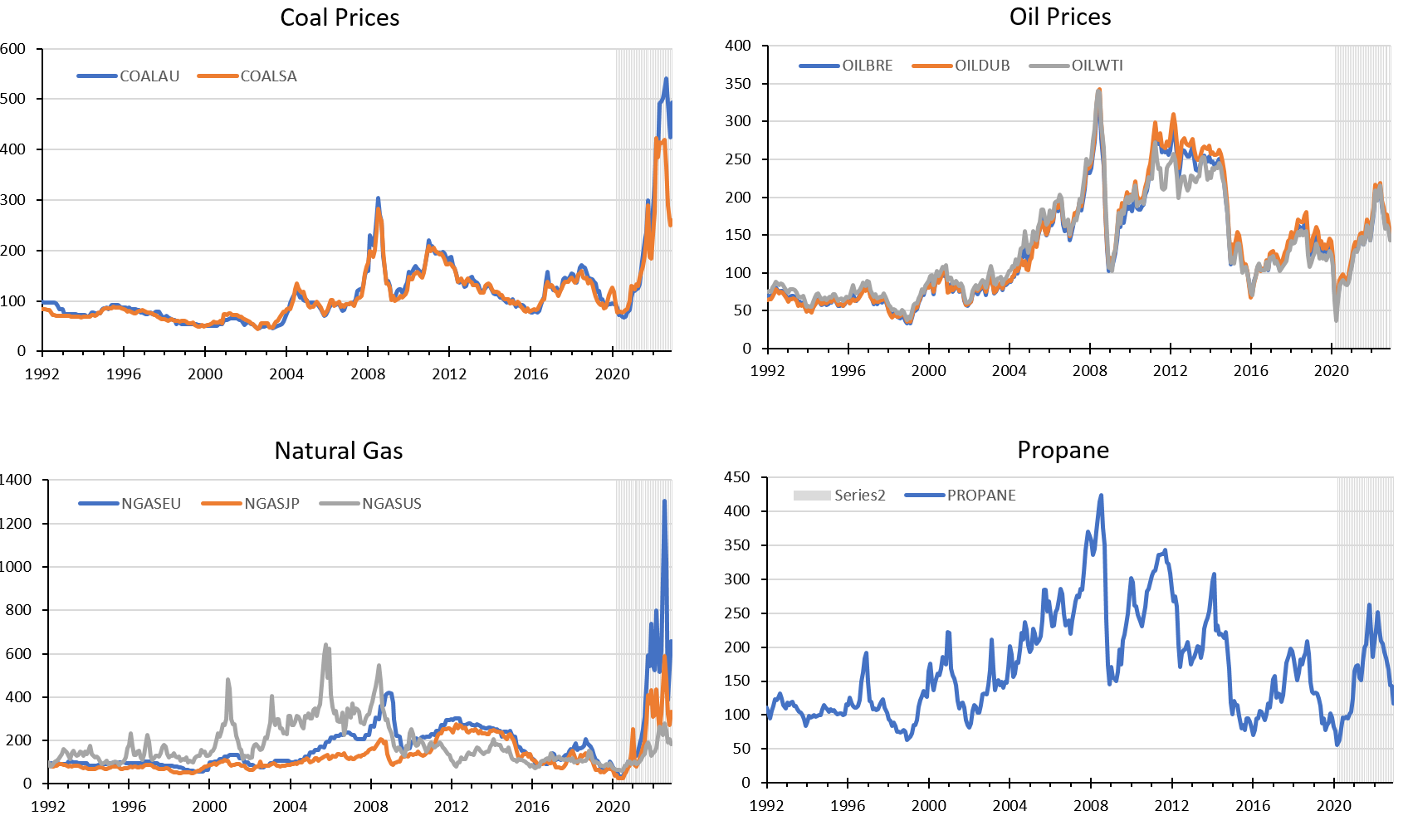}
\end{center}
\caption{Primary Energy Commodity Prices (Real terms based on IMF data, index 2016=100).} \label{Figure_1}
\end{figure}

We have applied M-CiSSA to the multivariate analysis of the monthly Primary Commodity Energy Prices included in the category ENERGY by the IMF. 
We have analyzed the sample that goes from January 1992 until December 2022 ($T=372$ observations).  ENERGY comprises a set of nine commodity prices: Australian and South African Coal (COALAU, COALSA); Brent, Dubai and West Texas Intermediate Crude Oil (OILBRE, OILDUB, OILWTI); European, Indonesian and US Natural Gas (NGASEU, NGASJP, NGASUS) and Propane (PROPANE). The original prices in US\$ have been deflated by the US Consumer Price Index to turn them into real terms\footnote{Consideration of the data directly in nominal terms yields the same conclusions. Results are available upon request.} and they have been transformed into index numbers (2016=100) for homogeneity. Figure \ref{Figure_1} shows their graphs.

Before applying CiSSA and M-CiSSA, we only need to choose one parameter, the window length $L$. Due to the monthly periodicity, the first consideration is that $L$ should be multiple of 12. On the other hand, $L$ must also be a multiple of the cycle periods to be analyzed as recommended by Golyandina and Zhigljavsky (2013). 
In this regard, as economic business cycles expand approximately every 8 years, we have chosen $L=12 \times 8 =96$ and built the trajectory matrix and the associated block circulant matrix comprising 96 blocks of size $9 \times 9$. 
After diagonalizing this matrix, each resulting block is associated with a frequency $w_k=\frac{k-1}{L}$ and within each block we can further diagonalize it and understand the fluctuations within each frequency. In particular, $k=1$ will correspond to the trend, $k=2$ to $6$ and their symmetric counterparts $96$ to $92$, to the business cycle.

Table \ref{Table_1} shows the accumulated contribution to the total variability of each frequency, computed as the ratio in (\ref{contribution}) but summing all the eigenvalues corresponding to the same block over the total sum of eigenvalues. The accumulated contribution shows that the trend is the most informative signal capturing 36.7\% of the total variability. Furthermore, 96-month (8 year) cycles explain 22.3\% of the total variability. 
The 48-month (4 year) cycles account for 16.0\% of the variability.
Therefore, by analyzing trends and 8 and 4-year cycles, we can explain 75.3\% of the variability of energy primary commodity prices. 

\begin{table}
\caption{Accumulated contribution of each frequency over total variability of the Primary Energy Commodity Prices.}  \label{Table_1}
\begin{center}
\begin{tabular}{c | c c c c c c c}
\hline
\textbf{k} & 1 & 2 and 96 & 3 and 95 & 4 and 94 & 5 and 93 & 6 and 92 \\
\hline
\textbf{Period} & inf & 96 & 48 & 32 & 24 & 19.2  \\
\textbf{Contribution} & 36.7 & 22.6 & 16.0 & 8.0 & 3.7 & 2.1  \\
\hline
\end{tabular}
\end{center}
\end{table}

\subsubsection{Long-run behaviour of Primary Energy Commodity Prices}
Information about the trend is represented by the first $9 \times 9$ block of the approximation to the cross spectral density. 
Remember that, according to the uniqueness theorem (Theorem \ref{Theorem_2}), the univariate estimation in CiSSA of an oscillatory component at each frequency is the sum of the 9 subcomponents associated with this frequency in the multivariate M-CiSSA setup. 
If the number of subcomponents required to pick up practically the total variability of the long-run behaviour of the individual series, is less than the total number of series, then there are common long-run trends. Disentangling what is common and what is idiosyncratic to each series at each frequency is one of the contributions of M-CiSSA.
Table \ref{Table_2} shows that the variability of the trend is mainly explained by the first three eigenvalues of this first block, as they account for 99.7\%. 

\begin{table}
\caption{Accumulated contribution of the subcomponents over the variability of each frequency of the Primary Energy Commodity Prices.}  \label{Table_2}
\begin{center}
\begin{tabular}{c c c c c}
\hline
\multirow{2}{*}{\textbf{k}} & \multirow{2}{*}{\textbf{Period}} & \multicolumn{3}{c}{\textbf{Subcomponents}} \\
& & 1 & 2 & 3 \\ \hline
1 & inf & 72.9 & 94.9 & 99.7 \\
2 and 96 & 96 & 66.8 & 85.5 & 98.1  \\
3 and 95 & 48 & 86.3 & 95.3 & 98.3  \\
\hline
\end{tabular}
\end{center}
\end{table}

\begin{figure}[ht]
\begin{center}
\includegraphics[width=5.9in]{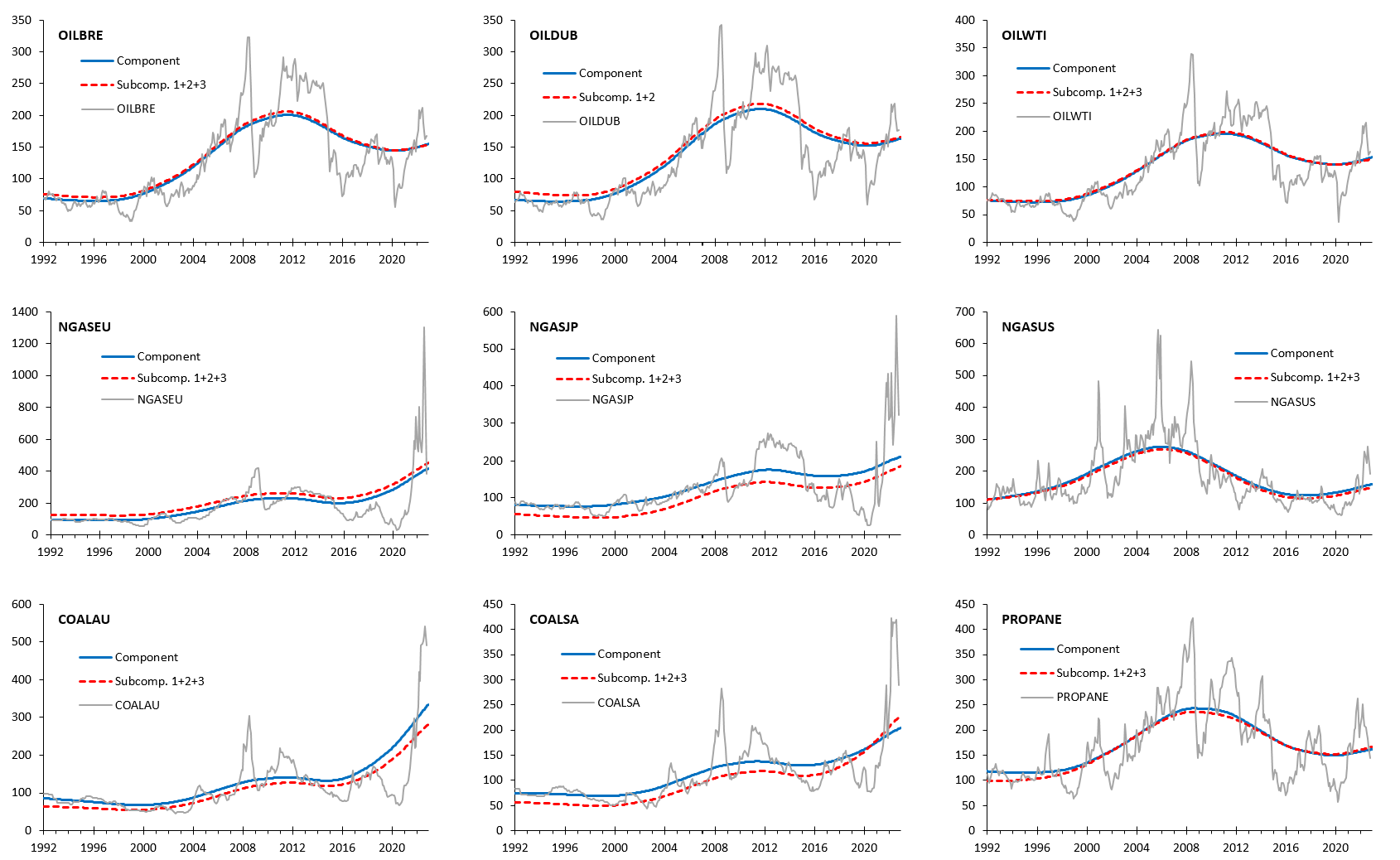}
\end{center}
\caption{Plot of the original series (in grey), the extracted univariate trend (in blue) and the sum of first, second and third trend subcomponents from the multivariate set-up (in red).}
\label{Figure_3}
\end{figure}

Figure \ref{Figure_3} shows the estimated trends for every energy commodity price and the reconstruction from the three first subcomponents. 
There is great similarity between the univariate trend and the multivariate one reconstructed by the 3 subcomponents, therefore understanding the flutctuations implied by the eigenvectors will characterize the long-run drivers of the 9 series. 

Looking at the individual series, we can see in the top panel of Table \ref{Table_4} that the 3 subcomponents explain almost all the variability of the trend of each variable, being enough to reconstruct the long run behaviour of each series (the minimum variability explained is 98.7\%  for NGASJP).  

\begin{table}
\caption{Accumulated contribution of the subcomponents of each series over the variability of the trend and 96-month and 48 cycles of the Primary Energy Commodity Prices.}  \label{Table_4}
\begin{center}
\resizebox{15.65cm}{!} {
\begin{tabular}{c c c c c c c c c c}
\hline
Subcomponent & COALAU & COALSA & OILBRE & OILDUB & OILWTI & NGASEU & NGASJP & NGASUS & PROPANE \\ \hline
\multicolumn{10}{c}{\textbf{Trend}} \\ \hline
1 & 66.9 & 77.1 & 95.0 & 95.0 & 94.9 & 87.8 & 88.5 & 6.2 & 74.6 \\
2 & 87.9 & 92.9 & 95.0 & 95.0 & 96.3 & 92.3 & 98.6 & 96.2 & 96.3 \\ 
3 & 99.5 & 99.4 & 99.9 & 99.9 & 99.9 & 99.9 & 98.7 & 99.9 & 99.0 \\ \hline
\multicolumn{10}{c}{\textbf{96-Month (8-year) Cycle}} \\ \hline
1 & 72.2 & 75.8 & 54.8 & 54.4 & 62.9 & 92.6 & 81.6 & 13.2 & 53.9 \\
2 & 73.0 & 76.0 & 74.0 & 75.5 & 83.0 & 95.8 & 91.9 & 72.1 & 93.1 \\ 
3 & 96.3 & 95.4 & 99.0 & 98.7 & 98.5 & 99.5 & 96.4 & 98.5 & 96.3 \\ \hline
\multicolumn{10}{c}{\textbf{48-Month (4-year) Cycle}} \\ \hline
1 & 95.3 & 95.1 & 69.0 & 68.6 & 69.4 & 96.4 & 97.0 & 35.3 & 60.0 \\
2 & 97.0 & 95.5 & 89.2 & 86.3 & 93.7 & 99.2 & 97.5 & 71.1 & 95.4 \\ 
3 & 97.1 & 95.9 & 97.3 & 96.6 & 97.4 & 99.5 & 98.3 & 99.6 & 97.1 \\
\hline
\end{tabular}
}
\end{center}
\end{table}

Additionally, the analysis of the eigenvectors also helps to understand the construction of the common forces of the trend.
From equation (\ref{UL}), the eigenvectors corresponding to the first block $k=1$ are just vectors of ones multiplied by their own constant, and they capture the changing level of the series. Each eigenvector can be subdivided in the subcomponents associated to each time series. Table \ref{Table_3} shows the relative weight of each variable\footnote{The relative weights are computed as $100\times$ the sum of the squares of the components of the eigenvector associated to each variable. Recall that eigenvectors have modulus 1.} in the first, second and third eigenvectors. 
The loadings\footnote{For brevity, the loadings are not shown but are available from the authors.} for the first eigenvector are all positive with a small contribution of NGASUS that becomes the main driver in the second eigenvector. Therefore, our first conclusion is that the long-run behaviour of NGASUS is decoupled from the rest of the energy prices. The third eigenvector gives positive weight to all oils and negative to coals and natural gases in Europe and the USA, therefore, it separates the oil market from the coal and natural gas markets.

\begin{table}
\caption{Relative weights of the main eigenvectors for the trend and the 96 and 48-month cycle of the Primary Energy Commodity Prices.}  \label{Table_3}
\begin{center}
\resizebox{15.65cm}{!} {
\begin{tabular}{c c c c c c c c c c}
\hline
Eigenvector & COALAU & COALSA & OILBRE & OILDUB & OILWTI & NGASEU & NGASJP & NGASUS & PROPANE \\ \hline
\multicolumn{10}{c}{\textbf{Trend}} \\ \hline
1 & 6.1 & 5.3 & 15.5 & 17.8 & 12.7 & 21.3 & 8.7 & 1.5 & 11.2 \\
2 & 6.4 & 3.6 & 0.0 & 0.0 & 0.6 & 3.6 & 3.3 & 70.8 & 11.8 \\ 
3 & 16.1 & 6.8  & 12.2  & 14.1 & 7.4  & 28.1  & 0.2 & 13.5 & 1.6 \\ \hline
\multicolumn{10}{c}{\textbf{96-month (8-year) Cycle}} \\ \hline
1 & 9.3 & 5.8 & 5.0 & 5.5 & 4.4 & 46.4 & 15.7 & 2.5 & 5.4 \\
2 & 0.4 & 0.1 & 7.2 & 8.9 & 5.8 & 6.8 & 8.1 & 46.6 & 16.1 \\ 
3 & 16.3 & 8.1 & 12.4 & 12.9 & 5.9 & 10.1 & 4.8 & 27.7 & 1.8 \\ 
\hline
\multicolumn{10}{c}{\textbf{48-month (4-year) Cycle}} \\ \hline
1 & 15.0 & 8.8 & 2.7 & 2.9 & 2.6 & 46.7 & 12.7 & 2.7 & 5.8 \\
2 & 2.5 & 0.3 & 7.7 & 7.3 & 8.7 & 13.2 & 0.7 & 26.5 & 33.0 \\ 
3 & 0.6 & 1.0 & 8.9 & 12.4 & 3.9 & 4.2 & 2.8 & 61.6 & 4.6 \\ 
\hline
\end{tabular}
}
\end{center}
\end{table}

\subsubsection{Cyclical behaviour of the Primary Energy Commodity Prices}

The second and third components in terms of relevance are the 96-month (8-year) and 48-month (4-year) cycles that account for the 22.6\% and 16.0\% of the total variability (Table \ref{Table_1}).

Regarding the 96-month cycle, it is represented by its corresponding $9 \times 9$ block for $k=2$ and $k=96$ of the cross power spectral density.
Table \ref{Table_2} shows that the variability within this frequency can be approximated (98.1\%) by the sum of the first 3 eigenvectors.
Table \ref{Table_4} shows the high approximation, over 95.4\% in all the series, of the sum of these 3 subcomponents.
Therefore, understanding the formation of the corresponding subcomponents allows to understand the common drivers of the 96-month cycle.

Table \ref{Table_3} shows the relative weights of each of the three eigenvectors. 
It can be seen that almost half of the total weight corresponds to NGASEU in the first eigenvector, while in  the second one, it is NGASUS the main driver. In the third eigenvector oil (specially OILBRE and OILDUB) and coal (COALAU and COALSA) gain more relevance compared to the previous subcomponents.

\begin{figure}[ht]
\begin{center}
\includegraphics[width=5in]{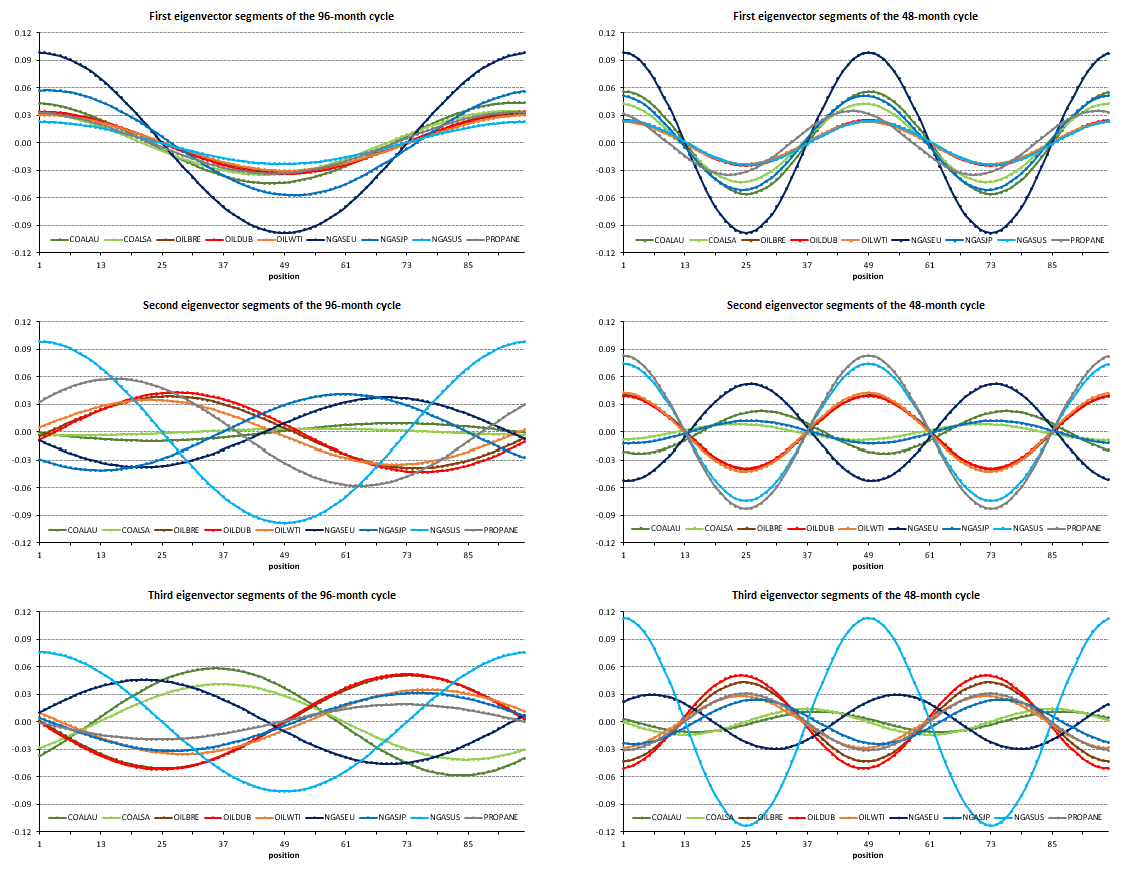}
\end{center}
\caption{Eigenvector segments for each variable of the 8-year and 4-year cycles of the Primary Energy Commodity Prices.} \label{Figure_4}
\end{figure}

The left panel of Figure \ref{Figure_4} shows the segments of the first, second and third eigenvectors corresponding to each commodity. 
The conclusions of Table \ref{Table_3} still hold at the look of the amplitude of the waves of the different eigenvector segments, however the analysis can be extended to understand the nature of the different cycles.
Regarding the first subcomponent (top-left graph) that explains (Table \ref{Table_2}) 66.8\% of the 96-month variability, we can see that all the segments share the same minima and maxima.
However, differences appear in the second and third subcomponents that jointly explain more than 30\% of the variability. The most striking effect is the decoupling in both graphs (left panel, middle and bottom graphs) of NGASUS (light blue line) as it has its own cycle in the second and third graphs, corresponding to additional subcomponents, different from the bulk of cycles that we see in the first graph (top panel, left column).

A similar analysis can be performed to understand the formation of the 4-year cycle.
Again, 3 subcomponents are needed in order to explain almost all the variability, $98.3\%$ (See Table \ref{Table_2}). And for the 4 year cycle associated to each variable, the sum of these 3 subcomponents explain more than $95.9\%$ (see Table \ref{Table_4}).
As it can be seen in Table \ref{Table_3}, the first subcomponent in the 4-year cycle is dominated by NGASEU, while the second one is mixture of NGASUS and PROPANE. Finally, the last subcomponent is mainly dominated only by NGASUS.
(Figure \ref{Figure_4}, right panel), we see that all prices are aligned for the first set of subvectors (top graph), showing NGASEU more amplitude in the cyclical 4-year wave for the first subcomponent, while  in the second subcomponent (right panel, middle graph) PROPANE and NGASUS are separated from NGASEU. Finally, the third subcomponent (right panel, bottom graph) NGASUS presents a specific movement opposed to the remaining variables. 

All in all, we have seen how 8 and 4-year cycles can be explained in a multivariate setup with M-CiSSA.

\section{Concluding Remarks}
We have introduced a new nonparametric methodology, M-CiSSA, that enables us to identify common fluctuations within a group of time series by frequency. It is also useful to deseasonalize and denoise the extracted signals.
M-CiSSA is based on the properties of block circulant matrices and its diagonalization that links the eigen-structure of this matrix and the  multivariate spectral density at different frequencies. 
Further diagonalization within each block of the spectral density matrix enables us to understand the multivariate information contained for each frequency.

We have proved that the uniqueness of each extracted component per frequency, that is, the extracted components in a univariate fashion coincide with the sum of all the subcomponents for each frequency extracted in a multivariate way. 
The added value of the multivariate approach is the ability to disentangle what is common from and what is idiosyncratic. This cannot be done using the univariate approach because there is no information regarding what is common. We depart from the abundant literature of dynamic factor models in several ways: firstly, we are able to separate common and idiosyncratic fluctuations by frequency; secondly, we can also discover phase shifts among the extracted signals for each series. Additionally, the multivariate approach is useful for further denoise the univariate extracted signals.  
We have illustrated the ability of M-CiSSA to separate different underlying signals through a synthetic example. We have also applied M-CiSSA to Primary Energy Commodity Prices.
On the one hand, understanding the formation of Primary Energy Commodity Prices is key for medium and long-term energy policy issues. 
In particular, we have discovered the nature of decoupling at different frequencies and that markets can have a very different behaviour in the medium and long run.  This is precisely one novelty of our approach: we can ascertain at what frequencies markets are decoupled. 
The comparison between the results in the univariate and multivariate cases also enables us to discover which prices are held or sustained by forces outside the global evolution of the markets. In particular, in the long run, we found that although coals are not decoupled, their prices are higher than suggested by the global forces of the markets. We also found that the long-run prices for natural gas in Europe exhibit a lower level than suggested, probably due to the European policy of installing liquid natural gas reserves.

 
Finally, our approach is quite general and can be applied to many disciplines involving time series analysis. Among other things, it can be used to separate long, medium, and short-run analysis, cyclical analysis, forecasting scenarios, deseasonalizing and denoising time series, and extracting the common from the idiosyncratic signals by frequency due to the ability of the procedure to extract signals at the desired frequencies specified by the user.


\begin{appendices}
\section{Theorems and Proofs}

\subsection{Proof of Theorem \ref{Theorem_1}}
${\widetilde{\mathbf{F}}}$ is a continuous matrix function and $\mathbf{C}_L\left({\widetilde{\mathbf{F}}}\right)$ is also a block Toeplitz matrix. Tilli (1998) holds that $\sigma_1\left(\mathbf{T}_L\left(\mathbf{F}\right)\right)\le\sigma_1\left(\mathbf{F}\right) < \infty$ and $\sigma_1\left(\mathbf{C}_L\left(\widetilde{\mathbf{F}}\right)\right)\le\sigma_1\left(\widetilde{\mathbf{F}}\right)<\infty \quad \forall L\in\mathbb{N}\:$, where $\sigma_1\left(\mathbf{A}\right)$ is the largest singular value of matrix $\mathbf{A}$.
We also must proof that 
$\underset{L\to \infty }{\mathop{\lim }}\,{{L}^{\tfrac{-1}{2}}}{{\left\| {{\mathbf{T}}_{L}}\left( \mathbf{F} \right)-{{\mathbf{C}}_{L}}\left( {{{\mathbf{\widetilde{F}}}}} \right) \right\|}_{F}}=0$. From (\ref{T_L}) and (\ref{omega_tilde}) we have that
\[\begin{split}
 & \tfrac{1}{L}\left\| {{\mathbf{T}}_{L}}\left( \mathbf{F} \right)-{{\mathbf{C}}_{L}}\left( {{{\mathbf{\widetilde{F}}}}} \right) \right\|_{F}^{2} \\ 
 & \quad \quad =\sum\limits_{k=1}^{L-1}{\tfrac{\left( L-k \right){{k}^{2}}}{{{L}^{3}}}\left( \left\| {{\mathbf{\Gamma }}_{k}}-{{\mathbf{\Gamma }}_{-L+k}} \right\|_{F}^{2}+\left\| {{\mathbf{\Gamma }}_{-k}}-{{\mathbf{\Gamma }}_{L-k}} \right\|_{F}^{2} \right)} \\ 
 & \quad \quad =\sum\limits_{r=1}^{M}{\sum\limits_{s=1}^{N}{\sum\limits_{k=1}^{L-1}{\tfrac{\left( L-k \right){{k}^{2}}}{{{L}^{3}}}\left( \left| {{\left[ {{\mathbf{\Gamma }}_{k}} \right]}_{r,s}}-{{\left[ {{\mathbf{\Gamma }}_{-L+k}} \right]}_{r,s}} \right|_{{}}^{2}+\left| {{\left[ {{\mathbf{\Gamma }}_{-k}} \right]}_{r,s}}-{{\left[ {{\mathbf{\Gamma }}_{L-k}} \right]}_{r,s}} \right|_{{}}^{2} \right)}\,.}} \\ 
\end{split}\]
As $\left[\mathbf{\Gamma}_k\right]_{r,s}\in\mathbb{C}$, we write $\left[\mathbf{\Gamma}_k\right]_{r,s}=a_{k,r,s}+i\cdot b_{k,r,s}$ where $a_{k,r,s}, b_{k,r,s}\in\mathbb{R}$. Therefore,
\[\begin{split}
 & \sum\limits_{k=1}^{L-1}{\left| {{\left[ {{\mathbf{\Gamma }}_{k}} \right]}_{r,s}}-{{\left[ {{\mathbf{\Gamma }}_{-L+k}} \right]}_{r,s}} \right|_{{}}^{2}}=\sum\limits_{k=1}^{L-1}{{{\left( {{a}_{k,r,s}}-{{a}_{-L+k,r,s}} \right)}^{2}}}+\sum\limits_{k=1}^{L-1}{{{\left( {{b}_{k,r,s}}-{{b}_{-L+k,r,s}} \right)}^{2}}} \\ 
 & \quad \quad =\sum\limits_{k=1}^{L-1}{a_{k,r,s}^{2}}+\sum\limits_{k=1}^{L-1}{a_{-L+k,r,s}^{2}}-2\sum\limits_{k=1}^{L-1}{a_{k,r,s}^{{}}a_{-L+k,r,s}^{{}}}\\
 & \quad \quad \quad \quad +\sum\limits_{k=1}^{L-1}{b_{k,r,s}^{2}}+\sum\limits_{k=1}^{L-1}{b_{-L+k,r,s}^{2}}-2\sum\limits_{k=1}^{L-1}{b_{k,r,s}^{{}}b_{-L+k,r,s}^{{}}} \\ 
 & \quad \quad =\sum\limits_{k=1}^{L-1}{\left| {{\left[ {{\mathbf{\Gamma }}_{k}} \right]}_{r,s}} \right|_{{}}^{2}}+\sum\limits_{k=1}^{L-1}{\left| {{\left[ {{\mathbf{\Gamma }}_{-L+k}} \right]}_{r,s}} \right|_{{}}^{2}}-2\sum\limits_{k=1}^{L-1}{a_{k,r,s}^{{}}a_{-L+k,r,s}^{{}}}-2\sum\limits_{k=1}^{L-1}{b_{k,r,s}^{{}}b_{-L+k,r,s}^{{}}}\ \ . \\ 
\end{split}\]
Considering that $2\vert xy\vert \le x^2+y^2 \quad \forall x,y\in\mathbb{R}$, we have that
\[\begin{split}
 & \sum\limits_{k=1}^{L-1}{\left| {{\left[ {{\mathbf{\Gamma }}_{k}} \right]}_{r,s}}-{{\left[ {{\mathbf{\Gamma }}_{-L+k}} \right]}_{r,s}} \right|_{{}}^{2}} \\ 
 & \quad \quad \le \sum\limits_{k=1}^{L-1}{\left| {{\left[ {{\mathbf{\Gamma }}_{k}} \right]}_{r,s}} \right|_{{}}^{2}}+\sum\limits_{k=1}^{L-1}{\left| {{\left[ {{\mathbf{\Gamma }}_{-L+k}} \right]}_{r,s}} \right|_{{}}^{2}}+\sum\limits_{k=1}^{L-1}{\left( a_{k,r,s}^{2}+a_{-L+k,r,s}^{2} \right)}+\sum\limits_{k=1}^{L-1}{\left( b_{k,r,s}^{2}+b_{-L+k,r,s}^{2} \right)} \\  & \quad \quad =2\left( \sum\limits_{k=1}^{L-1}{\left| {{\left[ {{\mathbf{\Gamma }}_{k}} \right]}_{r,s}} \right|_{{}}^{2}}+\sum\limits_{k=1}^{L-1}{\left| {{\left[ {{\mathbf{\Gamma }}_{-L+k}} \right]}_{r,s}} \right|_{{}}^{2}} \right)\ \ . \\ 
\end{split}\]
In the same way,
\[\sum\limits_{k=1}^{L-1}{\left| {{\left[ {{\mathbf{\Gamma }}_{-k}} \right]}_{r,s}}-{{\left[ {{\mathbf{\Gamma }}_{L-k}} \right]}_{r,s}} \right|_{{}}^{2}}\le 2\left( \sum\limits_{k=1}^{L-1}{\left| {{\left[ {{\mathbf{\Gamma }}_{-k}} \right]}_{r,s}} \right|_{{}}^{2}}+\sum\limits_{k=1}^{L-1}{\left| {{\left[ {{\mathbf{\Gamma }}_{L-k}} \right]}_{r,s}} \right|_{{}}^{2}} \right) .\]
As a consequence
\[\begin{split}
 & \sum\limits_{k=1}^{L-1}{\left( \left| {{\left[ {{\mathbf{\Gamma }}_{k}} \right]}_{r,s}}-{{\left[ {{\mathbf{\Gamma }}_{-L+k}} \right]}_{r,s}} \right|_{{}}^{2}+\left| {{\left[ {{\mathbf{\Gamma }}_{-k}} \right]}_{r,s}}-{{\left[ {{\mathbf{\Gamma }}_{L-k}} \right]}_{r,s}} \right|_{{}}^{2} \right)} \\ 
 & \quad \quad \le 2\left( \sum\limits_{k=1}^{L-1}{\left| {{\left[ {{\mathbf{\Gamma }}_{k}} \right]}_{r,s}} \right|_{{}}^{2}}+\sum\limits_{k=1}^{L-1}{\left| {{\left[ {{\mathbf{\Gamma }}_{-L+k}} \right]}_{r,s}} \right|_{{}}^{2}}+\sum\limits_{k=1}^{L-1}{\left| {{\left[ {{\mathbf{\Gamma }}_{-k}} \right]}_{r,s}} \right|_{{}}^{2}}+\sum\limits_{k=1}^{L-1}{\left| {{\left[ {{\mathbf{\Gamma }}_{L-k}} \right]}_{r,s}} \right|_{{}}^{2}} \right) \\ 
 & \quad \quad =4\left( \sum\limits_{k=-L+1}^{L-1}{\left| {{\left[ {{\mathbf{\Gamma }}_{k}} \right]}_{r,s}} \right|_{{}}^{2}}-\left| {{\left[ {{\mathbf{\Gamma }}_{0}} \right]}_{r,s}} \right|_{{}}^{2} \right)\le 4\sum\limits_{k=-L+1}^{L-1}{\left| {{\left[ {{\mathbf{\Gamma }}_{k}} \right]}_{r,s}} \right|_{{}}^{2}} \\ 
\end{split}\]
and, therefore,
\[\sum\limits_{k=1}^{L-1}{\tfrac{\left( L-k \right){{k}^{2}}}{{{L}^{3}}}\left( \left| {{\left[ {{\mathbf{\Gamma }}_{k}} \right]}_{r,s}}-{{\left[ {{\mathbf{\Gamma }}_{-L+k}} \right]}_{r,s}} \right|_{{}}^{2}+\left| {{\left[ {{\mathbf{\Gamma }}_{-k}} \right]}_{r,s}}-{{\left[ {{\mathbf{\Gamma }}_{L-k}} \right]}_{r,s}} \right|_{{}}^{2} \right)}\le 4\sum\limits_{k=-L+1}^{L-1}{\tfrac{\left( L-\left| k \right| \right){{k}^{2}}}{{{L}^{3}}}\left| {{\left[ {{\mathbf{\Gamma }}_{k}} \right]}_{r,s}} \right|_{{}}^{2}}.\]
Parseval's Theorem guaranties quadratic summability of $\left\{\left[\mathbf{\Gamma}_k\right]_{r,s}\right\}_{k\in\mathbb{Z}} \: \forall \: 1\le r\le M$ and $1\le s\le N$. So, given $\varepsilon>0$, you can choose $P>0$ big enough such that $\sum\limits_{k=P}^{\infty }{\left( \left| {{\left[ {{\mathbf{\Gamma }}_{k}} \right]}_{r,s}} \right|_{{}}^{2}+\left| {{\left[ {{\mathbf{\Gamma }}_{-k}} \right]}_{r,s}} \right|_{{}}^{2} \right)}\le \varepsilon$. Then,

\[\begin{split}
 & \underset{L\to \infty }{\mathop{\lim }}\,\sum\limits_{k=1}^{L-1}{\tfrac{\left( L-k \right){{k}^{2}}}{{{L}^{3}}}\left( \left| {{\left[ {{\mathbf{\Gamma }}_{k}} \right]}_{r,s}}-{{\left[ {{\mathbf{\Gamma }}_{-L+k}} \right]}_{r,s}} \right|_{{}}^{2}+\left| {{\left[ {{\mathbf{\Gamma }}_{-k}} \right]}_{r,s}}-{{\left[ {{\mathbf{\Gamma }}_{L-k}} \right]}_{r,s}} \right|_{{}}^{2} \right)}\\ 
 & \quad \quad \le 4\underset{L\to \infty }{\mathop{\lim }}\,\sum\limits_{k=-L+1}^{L-1}{\tfrac{\left( L-\left| k \right| \right){{k}^{2}}}{{{L}^{3}}}\left| {{\left[ {{\mathbf{\Gamma }}_{k}} \right]}_{r,s}} \right|_{{}}^{2}} \\ 
 & \quad \quad =4\underset{L\to \infty }{\mathop{\lim }}\,\left\{ \sum\limits_{k=-P+1}^{P-1}{\tfrac{\left( L-\left| k \right| \right){{k}^{2}}}{{{L}^{3}}}\left| {{\left[ {{\mathbf{\Gamma }}_{k}} \right]}_{r,s}} \right|_{{}}^{2}}+\sum\limits_{k=P}^{L-1}{\tfrac{\left( L-k \right){{k}^{2}}}{{{L}^{3}}}\left( \left| {{\left[ {{\mathbf{\Gamma }}_{k}} \right]}_{r,s}} \right|_{{}}^{2}+\left| {{\left[ {{\mathbf{\Gamma }}_{-k}} \right]}_{r,s}} \right|_{{}}^{2} \right)} \right\} \\ 
 & \quad \quad \le 4\underset{L\to \infty }{\mathop{\lim }}\,\sum\limits_{k=-P+1}^{P-1}{\tfrac{\left( L-\left| k \right| \right){{k}^{2}}}{{{L}^{3}}}\left| {{\left[ {{\mathbf{\Gamma }}_{k}} \right]}_{r,s}} \right|_{{}}^{2}}+4\sum\limits_{k=P}^{\infty }{\left( \left| {{\left[ {{\mathbf{\Gamma }}_{k}} \right]}_{r,s}} \right|_{{}}^{2}+\left| {{\left[ {{\mathbf{\Gamma }}_{-k}} \right]}_{r,s}} \right|_{{}}^{2} \right)}\le 0+4\varepsilon =4\varepsilon \ . \\ 
\end{split}\]
For all this,
\[\begin{split}
 & \underset{L\to \infty }{\mathop{\lim }}\,\tfrac{1}{L}\left\| {{\mathbf{T}}_{L}}\left( \mathbf{F} \right)-{{\mathbf{C}}_{L}}\left( {{{\mathbf{\widetilde{F}}}}} \right) \right\|_{F}^{2} \\ 
 & \quad \quad =\underset{L\to \infty }{\mathop{\lim }}\,\sum\limits_{r=1}^{M}{\sum\limits_{s=1}^{N}{\sum\limits_{k=1}^{L-1}{\tfrac{\left( L-k \right){{k}^{2}}}{{{L}^{3}}}\left( \left| {{\left[ {{\mathbf{\Gamma }}_{k}} \right]}_{r,s}}-{{\left[ {{\mathbf{\Gamma }}_{-L+k}} \right]}_{r,s}} \right|_{{}}^{2}+\left| {{\left[ {{\mathbf{\Gamma }}_{-k}} \right]}_{r,s}}-{{\left[ {{\mathbf{\Gamma }}_{L-k}} \right]}_{r,s}} \right|_{{}}^{2} \right)}\,}} \\ 
 & \quad \quad =\sum\limits_{r=1}^{M}{\sum\limits_{s=1}^{N}{\underset{L\to \infty }{\mathop{\lim }}\,\sum\limits_{k=1}^{L-1}{\tfrac{\left( L-k \right){{k}^{2}}}{{{L}^{3}}}\left( \left| {{\left[ {{\mathbf{\Gamma }}_{k}} \right]}_{r,s}}-{{\left[ {{\mathbf{\Gamma }}_{-L+k}} \right]}_{r,s}} \right|_{{}}^{2}+\left| {{\left[ {{\mathbf{\Gamma }}_{-k}} \right]}_{r,s}}-{{\left[ {{\mathbf{\Gamma }}_{L-k}} \right]}_{r,s}} \right|_{{}}^{2} \right)}\,}} \\ 
 & \quad \quad \le 4MN\varepsilon \,. \\ 
\end{split}\]
As $\varepsilon$ is any number,
\[\underset{L\to \infty }{\mathop{\lim }}\,{{L}^{\tfrac{-1}{2}}}{{\left\| {{\mathbf{T}}_{L}}\left( \mathbf{F} \right)-{{\mathbf{C}}_{L}}\left( {{{\mathbf{\widetilde{F}}}}} \right) \right\|}_{F}}=0\]
and, therefore, $\mathbf{T}_L\left(\mathbf{F}\right)\sim\mathbf{C}_L\left({\widetilde{\mathbf{F}}}\right)$ .

\subsection{Proof of Proposition \ref{Proposition_1}}
The solutions on the unit circle of the equation ${{z}^{n}}=1$ add up to zero and $\mathbf{V}$ is an unitary matrix. Therefore, we have that
\[\begin{array}{*{35}{l}}
 \sqrt{2}\mathcal{R}{{\text{ }\!\!'\!\!\text{ }}_{{{\mathbf{v}}_{i}}}}\sqrt{2}{{\mathcal{R}}_{{{\mathbf{v}}_{j}}}} & =\tfrac{1}{2}\left( {{\mathbf{v}}_{i}}+{{{\mathbf{\bar{v}}}}_{i}} \right)\text{ }\!\!'\!\!\text{ }\left( {{\mathbf{v}}_{j}}+{{{\mathbf{\bar{v}}}}_{j}} \right)=\tfrac{1}{2}\left[ \mathbf{v}{{\text{ }\!\!'\!\!\text{ }}_{i}}{{\mathbf{v}}_{j}}+\mathbf{v}_{i}^{*}{{\mathbf{v}}_{j}}+\overline{\left( \mathbf{v}_{i}^{*}{{\mathbf{v}}_{j}} \right)}+\overline{\left( \mathbf{v}{{\text{ }\!\!'\!\!\text{ }}_{i}}{{\mathbf{v}}_{j}} \right)} \right] \\
 {} & =\left\{ \begin{matrix}
 \tfrac{1}{2}\left( 0+0+0+0 \right)=0 & i\ne j \\
 \tfrac{1}{2}\left( 0+1+1+0 \right)=1 & i=j \\
\end{matrix} \right. \\
\end{array},\]
\[\begin{array}{*{35}{l}}
 \sqrt{2}\mathcal{I}{{\text{ }\!\!'\!\!\text{ }}_{{{\mathbf{v}}_{i}}}}\sqrt{2}{{\mathcal{I}}_{{{\mathbf{v}}_{j}}}} & =\tfrac{-1}{2}\left( {{\mathbf{v}}_{i}}-{{{\mathbf{\bar{v}}}}_{i}} \right)\text{ }\!\!'\!\!\text{ }\left( {{\mathbf{v}}_{j}}-{{{\mathbf{\bar{v}}}}_{j}} \right)=\tfrac{-1}{2}\left[ \mathbf{v}{{\text{ }\!\!'\!\!\text{ }}_{i}}{{\mathbf{v}}_{j}}-\mathbf{v}_{i}^{*}{{\mathbf{v}}_{j}}-\overline{\left( \mathbf{v}_{i}^{*}{{\mathbf{v}}_{j}} \right)}+\overline{\left( \mathbf{v}{{\text{ }\!\!'\!\!\text{ }}_{i}}{{\mathbf{v}}_{j}} \right)} \right] \\
 {} & =\left\{ \begin{matrix}
 \tfrac{-1}{2}\left( 0-0-0+0 \right)=0 & i\ne j \\
 \tfrac{-1}{2}\left( 0-1-1+0 \right)=1 & i=j \\
\end{matrix} \right. \\
\end{array}\]
\[\begin{array}{*{35}{l}}
 \sqrt{2}\mathcal{R}{{\text{ }\!\!'\!\!\text{ }}_{{{\mathbf{v}}_{i}}}}\sqrt{2}{{\mathcal{I}}_{{{\mathbf{v}}_{j}}}} & =\tfrac{-i}{2}\left( {{\mathbf{v}}_{i}}+{{{\mathbf{\bar{v}}}}_{i}} \right)\text{ }\!\!'\!\!\text{ }\left( {{\mathbf{v}}_{j}}-{{{\mathbf{\bar{v}}}}_{j}} \right)=\tfrac{-i}{2}\left[ \mathbf{v}{{\text{ }\!\!'\!\!\text{ }}_{i}}{{\mathbf{v}}_{j}}-\mathbf{v}_{i}^{*}{{\mathbf{v}}_{j}}+\overline{\left( \mathbf{v}_{i}^{*}{{\mathbf{v}}_{j}} \right)}-\overline{\left( \mathbf{v}{{\text{ }\!\!'\!\!\text{ }}_{i}}{{\mathbf{v}}_{j}} \right)} \right] \\
 {} & =\left\{ \begin{matrix}
 \tfrac{-i}{2}\left( 0-0+0-0 \right)=0 & i\ne j \\
 \tfrac{-i}{2}\left( 0-1+1-0 \right)=0 & i=j \\
\end{matrix} \right. \\
\end{array}.\]
Furthermore, since for the eigenvalue-eigenvector pair $\left({{\lambda }_{k,m}},{{\mathbf{v}}_{k,m}}\right)$ it is held ${{\mathbf{C}}_{L}}\left( \mathbf{F} \right){{\mathcal{R}}_{{{\mathbf{v}}_{k,m}}}}+\operatorname{i}{{\mathbf{C}}_{L}}\left( \mathbf{F} \right){{\mathcal{I}}_{{{\mathbf{v}}_{k,m}}}}={{\lambda }_{k,m}}{{\mathcal{R}}_{{{\mathbf{v}}_{k,m}}}}+\operatorname{i}{{\lambda }_{k,m}}{{\mathcal{I}}_{{{\mathbf{v}}_{k,m}}}}$, that is, ${{\mathbf{C}}_{L}}\left( \mathbf{F} \right){{\mathcal{R}}_{{{\mathbf{v}}_{k,m}}}}={{\lambda }_{k,m}}{{\mathcal{R}}_{{{\mathbf{v}}_{k,m}}}}$ and ${{\mathbf{C}}_{L}}\left( \mathbf{F} \right){{\mathcal{I}}_{{{\mathbf{v}}_{k,m}}}}={{\lambda }_{L+2-k,m}}{{\mathcal{I}}_{{{\mathbf{v}}_{k,m}}}}$ because ${{\lambda }_{k}}={{\lambda }_{L+2-k}}$, the proposition is proved.

\subsection{Proof of Theorem \ref{Theorem_2}}
 In M-CiSSA, the elementary matrix for the {\it m-th} subcomponent at frequency $\omega_k$ for the {\it i-th} series, using the unitary matrix $\mathbf{V}$ defined in (\ref{matrix_V}) and following the expression (\ref{X_kmi}), is given by
\begin{equation}
\mathbf{X}_{k,m}^{\left( i \right)}=\mathbf{v}_{k,m}^{\left( i \right)}\mathbf{v}_{k,m}^{*}\mathbf{X}\;.
\label{ X_kmi_short}
\end{equation}
The eigenvector ${{\mathbf{v}}_{k,m}}={{\mathbf{v}}_{\left( k-1 \right)M+m}}$ of the unitary matrix $\mathbf{V}$ can be expressed as
\begin{equation*}
{{\mathbf{v}}_{k,m}}=\left( {{\mathbf{U}}_{L}}\otimes {{\mathbf{I}}_{M}} \right)\left( {{\mathbf{1}}_{M,k}}\otimes {{\mathbf{e}}_{k,m}} \right)
\end{equation*}
what, together with formula (\ref{v_kmi}), transforms equation (\ref{ X_kmi_short}) into
\begin{equation}
\mathbf{X}_{k,m}^{\left( i \right)}=\left( {{\mathbf{I}}_{L}}\otimes \mathbf{1}_{M,i}^{T} \right)\left( {{\mathbf{U}}_{L}}\otimes {{\mathbf{I}}_{M}} \right)\left( {{\mathbf{1}}_{M,k}}\otimes {{\mathbf{e}}_{k,m}} \right){{\left( {{\mathbf{1}}_{M,k}}\otimes {{\mathbf{e}}_{k,m}} \right)}^{*}}{{\left( {{\mathbf{U}}_{L}}\otimes {{\mathbf{I}}_{M}} \right)}^{*}}\mathbf{X}\;.
\label{ X_kmi_large}
\end{equation}
Adding the previous equality (\ref{ X_kmi_large}) in {\it m}, we obtain that
\[\begin{array}{*{35}{l}}
\sum\limits_{m=1}^{M}{\mathbf{X}_{k,m}^{\left( i \right)}} & = & \left( {{\mathbf{I}}_{L}}\otimes \mathbf{1}_{M,i}^{T} \right)\left( {{\mathbf{U}}_{L}}\otimes {{\mathbf{I}}_{M}} \right)\left[ \sum\limits_{m=1}^{M}{\left( {{\mathbf{1}}_{M,k}}\otimes {{\mathbf{e}}_{k,m}} \right){{\left( {{\mathbf{1}}_{M,k}}\otimes {{\mathbf{e}}_{k,m}} \right)}^{*}}} \right]{{\left( {{\mathbf{U}}_{L}}\otimes {{\mathbf{I}}_{M}} \right)}^{*}}\mathbf{X} \\
{} & = & \left( {{\mathbf{I}}_{L}}\otimes \mathbf{1}_{M,i}^{T} \right)\left( {{\mathbf{U}}_{L}}\otimes {{\mathbf{I}}_{M}} \right)\operatorname{diag}\left( \mathbf{0},\cdots ,{{\mathbf{I}}_{M}},\cdots ,\mathbf{0} \right){{\left( {{\mathbf{U}}_{L}}\otimes {{\mathbf{I}}_{M}} \right)}^{*}}\mathbf{X} \\
{} & = & \mathbf{u}_{k}^{{}}\mathbf{u}_{k}^{*}\mathbf{X}_{{}}^{\left( i \right)} \\
{} & = & \mathbf{X}_{k}^{\left( i \right)} \\
\end{array}\]
where $\mathbf{X}_{{}}^{\left( i \right)}$ is the trajectory matrix for the {\it i-th} series and the identity matrix ${{\mathbf{I}}_{M}}$ occupies the {\it k-th} place in the block diagonal matrix $\operatorname{diag}\left( \mathbf{0},\cdots ,{{\mathbf{I}}_{M}},\cdots ,\mathbf{0} \right)$. Therefore, it is shown that $\sum\limits_{m=1}^{M}{\mathbf{X}_{k,m}^{\left( i \right)}}=\mathbf{X}_{k}^{\left( i \right)}$ and, as a consequence, the oscillatory component of any series for each frequency is unique.

\end{appendices}

\end{document}